\documentclass[article]{revtex4-1}

\usepackage[utf8]{inputenc}
\usepackage[english]{babel}
\usepackage[T1]{fontenc}
\expandafter\let\csname equation*\endcsname\relax

\expandafter\let\csname endequation*\endcsname\relax

\usepackage{amsmath,amssymb,latexsym}
\usepackage{hyperref}

\usepackage{tikz}
\usepackage{lipsum}

\usepackage{color}

\usepackage{pgfplots}
\pgfplotsset{compat=1.16}

\usepackage{qcircuit}
\usepackage{graphicx} 
\usepackage{caption,subcaption}

\newcommand{\ket}[1]{\left|#1\right\rangle}
\newcommand{\bra}[1]{\left\langle#1\right|}
\newcommand{\overlap}[2]{\left\langle#1|#2\right\rangle}

\newcommand{\0}{\ket{0}}
\renewcommand{\1}{\ket{1}}
\DeclareMathOperator{\tr}{tr}

\begin{document}
\title{Finding eigenvectors with a quantum variational algorithm}
\author{Juan Carlos Garc\'ia Escart\'in}
\email{juagar@tel.uva.es}  
\affiliation{Universidad de Valladolid, Dpto. Teor\'ia de la Se\~{n}al e Ing. Telem\'atica, Paseo Bel\'en n$^o$ 15, 47011 Valladolid, Spain}
\date{\today}

\begin{abstract}
This paper presents a hybrid variational quantum algorithm that finds a random eigenvector of a unitary matrix with a known quantum circuit. The algorithm is based on the SWAP test on trial states generated by a parametrized quantum circuit. The eigenvector is described by a compact set of classical parameters that can be used to reproduce the found approximation to the eigenstate on demand. This variational eigenvector finder can be adapted to solve the generalized eigenvalue problem, to find the eigenvectors of normal matrices and to perform quantum principal component analysis (QPCA) on unknown input mixed states. These algorithms can all be run with low depth quantum circuits, suitable for an efficient implementation on noisy intermediate scale quantum computers (NISQC) and, with some restrictions, on linear optical systems. In full scale quantum computers, where there might be optimization problems due to barren plateaus in larger systems, the proposed algorithms can be used as a primitive to boost known quantum algorithms. Limitations and potential applications are discussed.
\end{abstract}

\maketitle

\section{Introduction: Noisy intermediate scale quantum computers and hybrid algorithms.}
Scalable general purpose quantum computers could run algorithms that are more efficient than any classical alternative \cite{Sho97,Mon16}. However, at the present moment, the available technology is restricted to computers with a moderate number of qubits with a varying degree of noise. These computers are usually dubbed Noisy Intermediate Scale Quantum Computers, NISQC \cite{Pre18,BCK22}.

In this scenario, there has been a growing interest on hybrid quantum-classical algorithms \cite{MRB16,ECB21,CC22} where part of the work is shifted to a classical computer. In most cases there is a continuous feedback between the classical and the quantum computer, which has quantum circuits that are a function of a few parameters which are updated in the classical part of the algorithm according to the results of the quantum measurements from previous stages.

These algorithms are particularly useful in problems dealing with large state spaces which can be sampled quickly on a quantum computer without the need to explicitly write out the whole state, which has a size that grows exponentially with the size of the problem. 

A good example is quantum simulation, one of the most promising applications of quantum computing. For instance, in molecular simulation, the variational quantum eigensolver can efficiently search for a particular state in the Hilbert space of $n$ qubits, which grows exponentially with the size of the problem, and still achieve reasonable results even in the presence of noise \cite{MEA20,BBM20,TCC22}.

This approach based on using quantum circuits with a classical control also plays a key role in quantum machine learning \cite{MNK18,BLS19,SK19,SBS20,CAB21}.

In this paper, I present an application of these concepts to the search for an eigenvalue of a known unitary which can be generalized to different kinds of matrices with some restrictions. The algorithm is based on a modified version the SWAP test for state comparison and can be realized with circuits of a low depth (with few consecutive elementary gates). We only require a state preparation phase, a circuit for $U$ and one CNOT gate and one Hadamard gate for each qubit on which $U$ acts. At the end of the process, we obtain a list of classical parameters that can produce a good approximation to a random eigenvector of the desired matrix using a known classically controlled quantum circuit.

The same method can be adapted to perform Quantum Principal Component Analysis of an unknown mixed state, which can be cast as a problem of finding the eigenvectors of a Hermitian matrix. 

The paper starts with a brief reminder of the SWAP test for quantum state comparison and some hardware-efficient optimizations in Section \ref{SWAPtes}, followed by an explanation of the notation and methods of hybrid algorithms in Section \ref{hybrid}.

The core quantum variational eigenvector finder algorithm is presented in Section \ref{VariationalAlg}, including modifications to solve the generalized eigenvalue problem (Section \ref{generalized}) and to find the eigenstates of general normal matrices instead of just unitaries (Section \ref{normal}). Section \ref{PCA} shows how to adapt the variational eigenvector finder for Quantum Principal Component Analysis. 

Section \ref{optical} proposes an alternative implementation in optical systems instead of the usual quantum circuit model. Section \ref{comparison} compares the proposed quantum variational algorithms to similar variational algorithms for the same problems commenting the advantages of each method. Finally, Section \ref{discussion} discuses the strong and weak points of the variational approach for eigenvector determination and comments some potential applications in the short and long term.

\section{The SWAP test}
\label{SWAPtes}
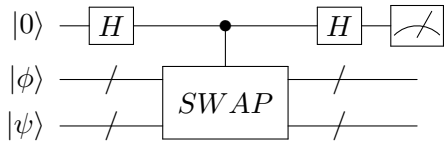
\begin{figure}[h]
\mbox{
\Qcircuit @C=1em @R=.7em {
& \lstick{\ket{0}} & \gate{H} & \ctrl{1} &\gate{H} & \meter \\
& \lstick{\ket{\phi}} & {/}\qw & \multigate{1}{SWAP}&\qw{/}&\qw\\
& \lstick{\ket{\psi}} & {/} \qw & \ghost{SWAP}&\qw{/}&\qw
}
}
\caption{Quantum circuit realizing the SWAP test. The crossed lines represent qubit registers with $n$ qubits.}\label{SWAPtestcircuit}
\end{figure}

While it is impossible to tell apart with certainty two quantum states which are not orthogonal, there are methods that can tell us if they are equal or not with a certain probability. The SWAP test \cite{BCW01,BBD97} permits a simple comparison. We start with two arbitrary $n$ qubit states $\ket{\psi}$ and $\ket{\phi}$ and an ancillary qubit initially in the state $\ket{0}$. The quantum circuit corresponding to the test is shown on Figure \ref{SWAPtestcircuit}.

The evolution in that circuit is

\begin{equation}
\begin{split}
&\ket{0}\ket{\phi}\ket{\psi}\overset{H}{\longrightarrow}\frac{\ket{0}+\ket{1}}{\sqrt{2}}\ket{\phi}\ket{\psi}\\ 
&\overset{\text{{\tiny CSWAP}}}{\longrightarrow}\frac{1}{\sqrt{2}}\left (\ket{0}\ket{\phi}\ket{\psi}+\ket{1}\ket{\psi}\ket{\phi}\right )\overset{\text{{H}}}{\longrightarrow}\\
&\ket{\Phi_{\text{out}}}=\frac{\ket{0}\left(\ket{\phi}\ket{\psi}+\ket{\psi}\ket{\phi}\right)+\ket{1}\left(\ket{\phi}\ket{\psi}-\ket{\psi}\ket{\phi}\right)}{2}.
\end{split}
\label{SWAPEvolution}
\end{equation}
If we measure the ancillary qubit after the last Hadamard gate, the probability of finding the state $\ket{0}$ is associated with the eigenvalue $1$ of the observable 
\begin{equation}
Z=\ket{0}\bra{0}-\ket{1}\bra{1}=P_{0}-P_{1}
\end{equation}
for the projectors to the $\ket{0}$ and $\ket{1}$ states $P_{0}$ and $P_{1}$. For the output state $\ket{\Phi_{\text{out}}}$ the probability of measuring the $\0$ state in the ancillary qubit is
\begin{equation}
\label{prob0}
P(0)=\bra{\Phi_{\text{out}}}P_0\ket{\Phi_{\text{out}}}=\frac{1+|\overlap{\psi}{\phi}|^2}{2}
\end{equation}
and the average value of the $Z$ observable for the first qubit becomes
\begin{equation}
\label{Z0}
\langle Z \rangle =|\overlap{\psi}{\phi}|^2.
\end{equation}

For two identical states, the SWAP operation
\begin{equation}
SWAP\ket{\psi}\ket{\phi}=\ket{\phi}\ket{\psi}
\end{equation}
does not change the state and the value of the control qubit is irrelevant. The two Hadamard gates on the ancillary qubit cancel and the whole circuit is equivalent to the identity. We can see from Eq. (\ref{SWAPEvolution}) that when both states are the same $\ket{\psi}=\ket{\phi}$ there is a destructive interference between the terms entangled to the ancillary qubit state $\ket{1}$. The result from the measurement is always $\ket{0}$ and whenever we get that result we say the SWAP test has been passed. 

However, if the states are different, each part of the uniform superposition of the ancillary qubit becomes entangled to a different state. In that case there is always a probability greater than zero of measuring the state $\ket{1}$. In a noiseless system, failing to pass the test shows with certainty the input states are different. The probability of finding a $\1$ ancillary qubit on measurement grows as the input states become more different. If we have $k$ copies of the same states and repeat the test $k$ times, even for very close states with an overlap $|\overlap{\psi}{\phi}|=1-\epsilon$ with $\epsilon\ll 1$, $P(0)=\frac{1+(1-\epsilon)^2}{2}\approx (1-\epsilon)$ and we have a probability of passing the test $P(0)^k\approx 1-k\epsilon$. With enough repetitions, if we never fail the test, we can be confident that the two input states are the same or, at least, very close.

As expected, global phases do not change the result of the test. Two inputs $\ket{\psi}$ and $e^{i\Phi}\ket{\psi}$, for an arbitrary phase $\Phi$, will act exactly in the same way under the test. This is the desired operation. While a larger circuit with a reference can distinguish between these two states, if these inputs were isolated states there is no physical measurement that could tell them apart.

The comparison in the SWAP test is also valid for general inputs with mixed states. We consider an ensemble given by the density matrix
\begin{equation}
\rho=\sum_{i}^{m}p_i\ket{\psi_i}\bra{\psi_i}
\end{equation}
where we have a statistical mixture of finding a pure state $\ket{\psi_i}$ with a probability $p_i$ for each. For two mixed states with density matrices $\rho$ and $\sigma$ the probability of passing the test becomes:
\begin{equation}
\label{prob0mix}
P(0)=\frac{1+\tr(\rho \sigma)}{2}
\end{equation}
where $\tr(\rho \sigma)$ replaces $|\overlap{\psi}{\phi}|^2$ \cite{KMY03}. When both states are pure states with $\rho=\ket{\psi}\bra{\psi}$ and $\sigma=\ket{\phi}\bra{\phi}$ we recover the original overlap.

\subsection{The destructive SWAP test}
The depth of the SWAP test circuit can be reduced if we notice that, after the test, the original inputs become entangled and they are discarded. The total number of elementary gates needed to perform a SWAP test can be reduced to $2n$ for $n$ qubits. Most importantly, we only need two stages. 

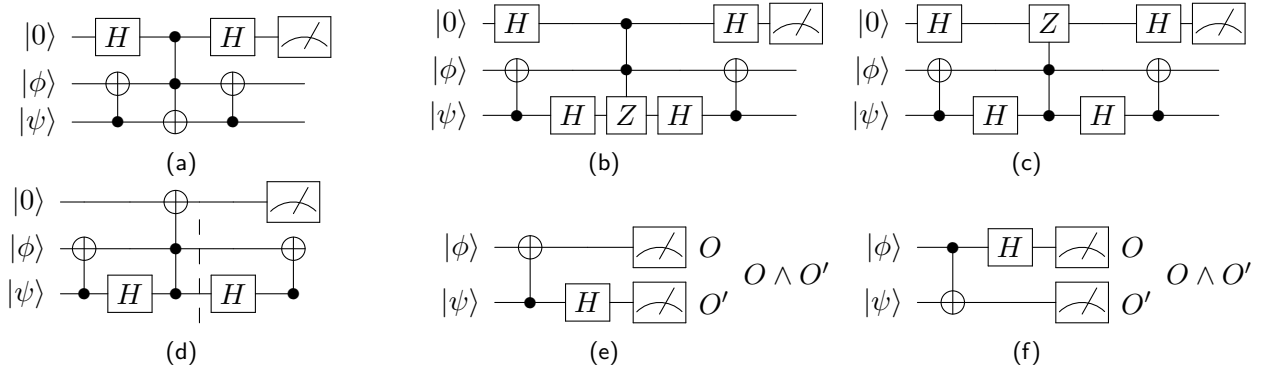
\begin{figure}
\centering 
\begin{subfigure}{0.25\textwidth}
 \makebox{
\Qcircuit @C=0.8em @R=.5em {
& \lstick{\ket{0}} & \gate{H} & \ctrl{1}  &\gate{H} & \meter \\
& \lstick{\ket{\phi}} & \targ & \ctrl{1} &\targ&\qw\\
& \lstick{\ket{\psi}} & \ctrl{-1} & \targ &\ctrl{-1}&\qw
}
}
\caption{}
\label{1qSWAPa} 
\end{subfigure}
\hfill
\begin{subfigure}{0.25\textwidth}
\makebox{
\Qcircuit @C=0.4em @R=.5em {
& \lstick{\ket{0}}    & \gate{H}  & \qw      & \ctrl{1}  & \qw      &\gate{H} & \meter \\
& \lstick{\ket{\phi}} & \targ     & \qw      & \ctrl{1}  & \qw      &\targ&\qw\\
& \lstick{\ket{\psi}} & \ctrl{-1} & \gate{H} & \gate{Z}  & \gate{H} &\ctrl{-1}&\qw
}
}
\caption{}
\label{1qSWAPb}
\end{subfigure}
\hfill
\begin{subfigure}{0.25\textwidth}
\makebox{
\Qcircuit @C=0.4em @R=.5em {
& \lstick{\ket{0}}    & \gate{H}  & \qw      & \gate{Z}  & \qw      &\gate{H} & \meter \\
& \lstick{\ket{\phi}} & \targ     & \qw      & \ctrl{-1}  & \qw      &\targ&\qw\\
& \lstick{\ket{\psi}} & \ctrl{-1} & \gate{H} & \ctrl{-1}  & \gate{H} &\ctrl{-1}&\qw
}
}
\caption{}
\label{1qSWAPc}
\end{subfigure}
\hfill

\begin{subfigure}{0.25\textwidth}
\makebox{
\Qcircuit @C=0.4em @R=.5em {
& \lstick{\ket{0}}    & \qw  & \qw      & \targ  & \qw      &\qw & \meter \\
& \lstick{\ket{\phi}} & \targ     & \qw      & \ctrl{-1}  &  \qw  \ar@{--}[]+<0em,1em>;[d]+<0em,-1em> &\qw &\targ\\
& \lstick{\ket{\psi}} & \ctrl{-1} & \gate{H} & \ctrl{-1}   &\qw & \gate{H} &\ctrl{-1}
}
}
\caption{}
\label{1qSWAPd}
\end{subfigure}
\hfill
\begin{subfigure}{0.25\textwidth}
\makebox{
\Qcircuit @C=0.8em @R=.5em {
& \lstick{\ket{\phi}} & \targ     & \qw      & \meter &\rstick{\hspace{-2ex}O}&\dstick{\phantom{aaaaaaa}{\small O \land O'}}\\
& \lstick{\ket{\psi}} & \ctrl{-1} & \gate{H} &\meter &\rstick{\hspace{-2ex}O'}
}
}
\caption{}
\label{1qSWAPe}
\end{subfigure}
\hfill
\begin{subfigure}{0.25\textwidth}
\makebox{
\Qcircuit @C=0.8em @R=.5em {
& \lstick{\ket{\phi}} & \ctrl{1}  & \gate{H}    & \meter &\rstick{\hspace{-2ex}O} &\dstick{\phantom{aaaaaaa}{\small O \land O'}}\\
& \lstick{\ket{\psi}} & \targ     & \qw         & \meter &\rstick{\hspace{-2ex}O'}
}
}
\caption{}
\label{1qSWAPf}
\end{subfigure}
\hfill
\caption{Equivalent circuits for the SWAP test on single qubit states.}
\label{destructiveSWAP}
\end{figure}

This is the destructive SWAP test explained in \cite{GC13}, which was later rediscovered by an AI driven search \cite{CSS18}. We can check the test is valid starting from the SWAP circuit for two qubits shown in Figure \ref{1qSWAPa}. The SWAP operation is performed with three CNOT gates. A sequence of 3 CNOT gates can perform classical XOR swapping \cite{War12} for two bits in the computational basis $x, y \in \mathbb{F}_2$ so that
\begin{align}
&\ket{x}\ket{y}\overset{\text{\tiny{CNOT(2,1)}}}{\longrightarrow}\ket{x\oplus y}\ket{y}\overset{\text{\tiny{CNOT(1,2)}}}{\longrightarrow}
\ket{x\oplus y}\ket{y\oplus x \oplus y}=\ket{x\oplus y}\ket{x}&\nonumber \\
&\overset{\text{\tiny{CNOT(2,1)}}}{\longrightarrow}\ket{x\oplus y \oplus x}\ket{x}=\ket{y}\ket{x}&
\end{align}
where CNOT($i,j$) is controlled by the $i$th qubit and has the $j$th qubit as target. The result follows from two properties the XOR operation $\oplus$ (equivalent to modulo 2 addition): $x\oplus x=0$ and $x\oplus 0=x$.

If the middle CNOT is replaced by a Toffoli gate controlled by the ancillary qubit in the SWAP test, we have a controlled SWAP operation. When the ancilla has a state $\1$ the middle Toffoli gate is equivalent to a CNOT gate on the inputs and we recover the 3 CNOTs of the SWAP circuit. When it is $\0$ the two consecutive CNOT gates cancel each other and we have the identity. 

The NOT operation can be decomposed as $X=HZH$ and we can convert CNOT gates into CZ gates surrounded by two Hadamard gates. The Toffoli gate becomes a doubly controlled Z gate with operation $CCZ\ket{x}\ket{y}\ket{z}=(-1)^{x\cdot y \cdot z}\ket{x}\ket{y}\ket{z}$ where the sign change only happens if all the qubits have a 1. In that sense, any of the qubits can be considered to be the target. Putting all together, we can find a series of equivalent circuits shown in Figs. \ref{1qSWAPb}-\ref{1qSWAPd}.
        
Now, starting from Figure \ref{1qSWAPd}, we can notice that changes on the input qubits after the Toffoli gate do no affect the state of the ancillary qubit. We might just as well measure both qubits right after the Toffoli. Finally, there is a further simplification if we notice that the ancillary qubit will result in a $\0$ unless both control qubits from the inputs are in state $\1$. The result of measuring the two input qubits at this point perfectly determines the state of the ancillary qubit, which can be removed. If we measure in the computational basis (observable $Z$), we can associate a measurement finding a $\0$ state to a bit $0$ ($Z$'s eigenvalue 1) and finding state $\1$ to a bit $1$ ($Z$'s eigenvalue $-1$). The logical AND of the measurement results reproduces the result of measuring the ancillary qubit. A measurement on the input qubits and classical computation (the AND operation) return the same statistics as measuring the $Z$ observable for the ancilla. 

The same reasoning can be applied to multiple qubits. In that case, the SWAP operation has the same decomposition for each pair of qubits of the multiqubit state. We group the first qubit of $\ket{\phi}$, $q_\phi^1$, with the first qubit of $\ket{\psi}$, $q_\psi^1$, and continue forming all the $n$ $(q_\phi^i,q_\psi^i)$ pairs for $1\leq i \leq n$. For $n$ qubit states, the simplified circuit in Figure \ref{1qSWAPd} would have the same CNOT and Hadamard gates in the corresponding qubit pairs and a Toffoli gate controlled by them and acting on the ancillary qubit. There is a total of $n$ Toffolis that change the state of the input from $\0$ to $\1$ or from $\1$ to $\0$ when both controls are $\1$. The final value depends on the parity of the bitwise AND, $\land$, of the strings coming from measuring the $n$ qubits of each input state. An even number of ones means the output state in the ancillary qubit would have been $\0$ (a pass). An odd number of ones corresponds to failing the test. 

In terms of the $Z$ observable of the ancillary qubit, the measurement result is exactly the same as the product of the measurement results for the $CZ$ observable on each pair. For two qubits, the $CZ$ operation has eigenvectors $\ket{00}, \ket{01}, \ket{10}$ associated to eigenvalue $1$ and an eigenvector $\ket{11}$ associated to eigenvalue $-1$. The product of these $CZ$ observables has the same value as a measurement of the observable $Z$ in the SWAP test. Considering all the qubits, the observable with the result of the destructive SWAP test is $CZ^{\otimes n}$ if we write our inputs in the order $q_\phi^1 q_\psi^1q_\phi^2 q_\psi^2\cdots q_\phi^n q_\psi^n$. Finding a $1$ eigenvalue corresponds to passing the test. Measuring eigenvalue $-1$ indicates a failure.

\section{Hybrid algorithms: Variational methods}
\label{hybrid}
Physical quantum computers are difficult to build. They have a high cost of operation when compared to classical computers and, in their current implementations, are still too prone to decoherence to be able to carry out long calculations. For all these reasons, whenever possible, we would like to replace the quantum parts of the algorithm by classical computation. For instance, in Shor's algorithm for integer factorization \cite{Sho97}, the only part that needs a quantum computer is the order finding subroutine. As shown by Eker\aa{ }\cite{Eke21}, a classical computer can perform a longer preprocessing stage and the quantum computer can be reserved for a single use of the non-classical routine for order finding. Even this simplified form of factoring still requires a number of qubits and consecutive stages beyond the capabilities of current noisy quantum computers.  

The first promising results of real-world quantum computers come from hybrid algorithms where the quantum and the classical stages feed each other. One particularly successful family of algorithms are variational methods taking advantage of the native capacity of a quantum computer to describe states in a large space without needing to store an exponential amount of data \cite{MEA20,TCC22}. 

For instance, if we want to compute the ground state energy $E_g$ of a particular Hamiltonian, $H$, classical variational methods propose a series of trial states. For any state $\ket{\psi}$, the expected energy value $\bra{\psi}H\ket{\psi}\geq E_g$ gives an upper bound to the minimal energy of the ground state. If we choose with care the states for which we measure the energy, we can approximate $E_g$ with good precision. Quantum methods offer a compact way to perform variational algorithms. We no longer need to write down the whole state vector of the ground state, which can have a number of complex elements growing exponentially with the size of the problem. For instance, a simple system with $n$ electrons in separate sites can have $2^n$ spin configurations.

Most variational hybrid algorithms can be described using four main blocks. We will describe them in parallel with an example application for ground state estimation for a Hamiltonian made of a combination of Pauli operators (Figure \ref{Variational}). The four blocks are:

\begin{itemize}
\item A parametrized quantum circuit $\mathbf{P}(\vec{\theta})$:

The gates in the circuit depend on a list of classical parameters $\vec{\theta}=(\theta_1,\ldots,\theta_k)$ to prepare a trial state, or \emph{ansatz}, $|\psi(\vec{\theta})\rangle$. We assume a fixed initial state, usually with the $n$ qubits initialized to $\0$ so that  $\mathbf{P}(\vec{\theta})\ket{0}^{\otimes n}=|\psi(\vec{\theta})\rangle$.

In the example of Figure \ref{Variational}, the classical parameters appear in a series of rotation gates $RX$ and $RY$ which rotate a single qubit with respect to the $X$ and $Y$ axis of the Bloch sphere, respectively, by an angle determined by a classical parameter. There are also two CNOT gates that introduce entanglement in the final state. This circuit follows the approach of what is called a Hardware Efficient Ansatz \cite{KMT17}, where the gates are selected to give the shallower possible circuit that still can approximate the desired state.

\item A processing circuit:

Depending on the desired target, we need to act on the ansatz in different ways. The measurement stage must return values related to an objective function scoring how close we are to our goal. 

The example in Figure \ref{Variational} shows a typical variational simulation scenario where the gates perform a change of basis so that the final measurements correspond to one term of the desired Hamiltonian in a particular encoding. In this case, the $H$ gate in the middle qubit and the identity in the rest make the final measurement in the computational basis give the exact same results as measuring the $Z\otimes X \otimes Z$ observable for the generated \emph{ansatz} $|\psi(\vec{\theta})\rangle$.

\begin{figure}[h]
\mbox{
\Qcircuit @C=1em @R=.7em {
\push{} & \push{} & \push{} & \push{} & \push{} & \push{} & \push{} & \push{} & \push{} & \push{}  & \push{} & \push{} & \push{} & \push{} & \push{} & \push{} \\
\lstick{\ket{0}} &  \qw  \gategroup{1}{2}{5}{7}{.7em}{--} &  \gate{RX(\theta_1)} & \ctrl{1}  &\qw      & \gate{RY(\theta_4)}& \qw  &\qw \ar@{--}[]+<1em,1.5em>;[d]+<1em,-3em>    & \qw  &  \qw & \qw& \qw   & \meter\\
\lstick{\ket{0}} &   \qw &  \gate{RX(\theta_2)} & \targ     & \ctrl{1}    & \gate{RY(\theta_5)}  &\qw   & \qw   & \qw & \qw \gategroup{1}{11}{5}{14}{.7em}{--}     &\qw  &\gate{H} & \meter  & \push{} &\rstick{f(\vec{\theta})} \\
\lstick{\ket{0}} &   \qw &  \gate{RX(\theta_2)} & \qw       & \targ    & \gate{RY(\theta_6)}   & \qw &  \qw  &\qw &  \qw  & & \qw \qw& \meter \\
\push{} & \push{} & \push{} & \push{} & \push{} & \push{} & \push{} & \push{} & \push{} & \push{} & \push{} & \push{} & \push{} & \push{} & \push{}& \push{} & \push{} & \push{}  \\
\push{} & \push{} & \push{} & \push{} & \push{U(\,\vec{\theta}\,)} & \push{} & \push{} & \push{} &\ustick{\scriptstyle{|\psi(\vec{\theta})\rangle}} & \push{} & \push{} & \push{} & \push{} & \push{} & \push{} \\
}
}
\caption{Example for three qubits $\langle\psi(\vec{\theta})|Z\otimes X\otimes Z|\psi(\vec{\theta})\rangle$.}\label{Variational}
\end{figure}
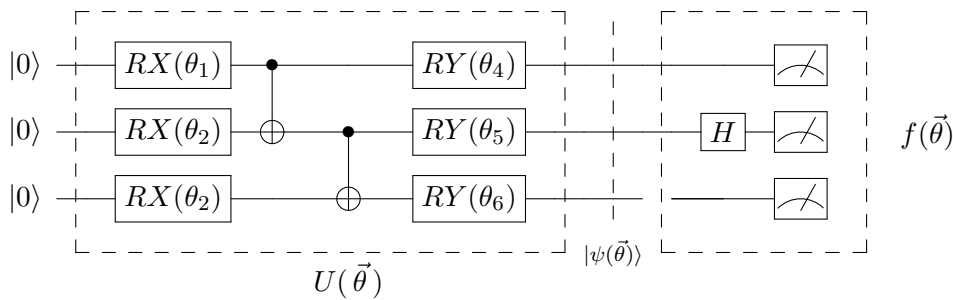

\item Measurement:

We usually consider measurement in the computational basis $\{\0,\1\}$ for each qubit. As a single measurement is not enough to estimate the desired averages, we need to prepare multiple copies of the same trial state using the exact same classical parameters and processing circuit so that we can get meaningful statistics. This stage has been considered together with the processing circuit in the example.

\item Classical processing and feedback:

After a few repetitions, the statistics from the measurement are used to compute an \emph{objective function} $f(\vec{\theta})$ which takes a value which depends on the chosen classical parameters and is related to the problem we want to solve. This dependence is, in general, quite complex. Using a quantum circuit we reduce an explicit description of the problem to measuring a small experimental setup. 

The global process is a loop that takes the estimated value of the objective function $f(\vec{\theta})$ at each iteration as a guide to tweak the classical parameters $\vec{\theta}$. In most cases, the problem is posed as a minimization problem where we use classical optimization methods (such as nonlinear optimization or gradient descent methods) changing the input parameters at stage $k$ to new values $\vec{\theta}_k$ until we find a stable value $f(\vec{\theta}_s)$ at iteration $s$ which is assumed to be the minimum.

In the example, the expected value $\langle\psi(\vec{\theta})|Z\otimes X\otimes Z|\psi(\vec{\theta})\rangle$, which approximates the energy of $|\psi(\vec{\theta})\rangle$ for the simulated Hamiltonian, is the objective function to be minimized. The classical part starts by choosing a random initial list of parameters $\vec{\theta}_1$ that gives a first value $f(\vec{\theta}_1)$. The chosen optimization method then generates a new set of classical parameters $\vec{\theta}_2$ that are fed into the parametrized quantum circuit. After we complete the second round of measurements, the new value of the objective function $f(\vec{\theta}_2)$ is compared to the previous one. The changes in the parameters that lead to a smaller final value are kept and the ones that increase the objective function are modified until we can no longer find a better solution. This part of the hybrid algorithm draws heavily on classical optimization and there are multiple available software libraries, like Python's scikit-learn \cite{PVG11}.
\end{itemize}

Notice that, in many cases, we have no guarantee that the algorithm will succeed. Still, we can provide educated guesses for the solution, similar to what happens in many classical machine learning methods. 

The main advantage of this family of hybrid algorithms is the possibility to probe a large state space with a quantum circuit that grows linearly instead of exponentially with the size of the system. This is particularly interesting in the simulation of quantum systems, where the states of the simulated system can be easily mapped into the quantum computer. For instance, chains of $n$ fermions with spin $1/2$ or $-1/2$ can be directly represented by $n$ qubits in the state $\0$ or $\1$ and we can directly explore the Hilbert space of dimension $2^n$ with a compact system and study, among others, phase transitions in the Ising model \cite{SJG22}. A good review of existing applications can be found in \cite{TCC22}.

\section{A variational method for the eigenvectors of unitary matrices}
\label{VariationalAlg}
The SWAP test gives a suitable objective function for a hybrid variational algorithm that finds the eigenstates $\ket{e_i}$ of a unitary evolution $U$ such that $U\ket{e_i}=\lambda_i\ket{e_i}$. The method can be generalized to multiple scenarios.

For an $N\times N$ unitary $U$, we can find $N$ orthogonal eigenvectors $\ket{e_i}$ which form an orthonormal basis and their corresponding eigenvalues $\lambda_i=e^{i\phi_i}$ are complex numbers of modulus 1.

\begin{figure}[h]
\mbox{
\Qcircuit @C=1em @R=.7em {
\lstick{\ket{0}} &   \qw         &  \multigate{3}{\mathbf{P}( \vec{\theta}\,)}  	& \qw  		&  \multigate{3}{\,U\,}	& \qw 		& \ctrl{5}	& \qw  		& \qw 		& \dots & & \qw  & \qw &\gate{H}  & \meter &\rstick{O_1}\\
\lstick{\ket{0}} &   \qw         & \ghost{\mathbf{P}( \vec{\theta}\,)} 		& \qw 		&  \ghost{\,U\,}  	& \qw 		& \qw  		& \ctrl{5}	& \qw 		& \dots & & \qw  &\qw  &\gate{H} & \meter &\rstick{O_2}\\
\push{}          & \push{\vdots} &  \push{}  				& \push{\vdots} & \push{} 		& \push{} 	& \push{}	& \push{}	& \push{}       & \push{\ddots} 	& \push{} & \push{}& \push{} & \push{} &  \push{\vdots}\\
\lstick{\ket{0}} &   \qw         & \ghost{\mathbf{P}( \vec{\theta}\,)}   		& \qw 		&   \ghost{\,U\,}  	& \qw 		&  \qw  	&  \qw 		&\qw            & \dots &  & \ctrl{5}  &\qw  &\gate{H} & \meter &\rstick{O_n}\\
\push{}          & \push{}       & \push{} 				&  \push{} 	& \push{} 		& \\
\lstick{\ket{0}} & \qw           &  \multigate{3}{\mathbf{P}( \vec{\theta}\,)}  	& \qw 		& \qw 			&\qw  		& \targ  	& \qw  		& \qw 		& \dots & & \qw  & \qw  &\gate{H} & \meter &\rstick{O_1'}\\
\lstick{\ket{0}} & \qw           & \ghost{\mathbf{P}( \vec{\theta}\,)}  		& \qw 		&  \qw 			& \qw  		& \qw  		& \targ 	&\qw 		& \dots & & \qw  & \qw   &\gate{H} & \meter &\rstick{O_2'}\\
\push{}          & \push{\vdots} &  \push{} 				& \push{}	& \push{}		& \push{\vdots} & \push{} 	& \push{}	& \push{} & \push{\ddots} & \push{} & \push{}& \push{} & \push{} &  \push{\vdots}\\
\lstick{\ket{0}} & \qw           & \ghost{\mathbf{P}( \vec{\theta}\,)}  		& \qw 		&  \qw 			& \qw 		& \qw 		&  \qw 	 	& \qw           & \dots & &  \targ& \qw   &\gate{H}& \meter &\rstick{O_n'}\\
}
}
\caption{Variational quantum circuit for the eigenvectors of $U$.}\label{VarU}
\end{figure}

The circuit for the quantum part of the algorithm is shown in Figure \ref{VarU}. We use the same parametrized circuit $\mathbf{P}(\vec{\theta})$ with the same classical parameters twice to generate a trial state $|\psi(\vec{\theta})\rangle|\psi(\vec{\theta})\rangle$, which is transformed by the unitary into $|\psi(\vec{\theta})\rangle U|\psi(\vec{\theta})\rangle$ and then proceeds to a destructive SWAP test circuit.  

The objective function in the algorithm comes from the statistics of the SWAP test and can be written as
\begin{equation}
f(\vec{\theta})=P(0)-P(1)=|\langle \psi(\vec{\theta})|U|\psi(\vec{\theta})\rangle|^2.
\end{equation}
We approximate the probabilities of passing and failing the SWAP test, $P(0)$ and $P(1)$ respectively, from counting how many times we get each outcome and dividing by the total number of times we run the test. The $0$ and $1$ results correspond to the parity of the bitwise AND of the measurement results in the computational basis (assuming $0$ for $\0$ and $1$ for $\1$) and they have the same statistics that would appear if we had measured the ancillary qubit in the complete SWAP test. The evaluation of the objective function is reduced to computing the expected value of the $Z$ observable for the ancillary qubit, albeit in a roundabout way to reduce the final number of gates.

The objective function is an inner product of normalized states, with $f(\vec{\theta})\leq 1$, and we can see that the maximum is obtained only if the trial state is an eigenvector of $U$, $|\psi(\vec{\theta}) \rangle=\ket{e_i}$. Then
\begin{equation}
f(\vec{\theta})=|\bra{e_i}U\ket{e_i}|^2=|\bra{e_i}\lambda_i\ket{e_i}|^2=1.
\end{equation}

Most optimization methods and software libraries work by minimizing a function. For these we can consider:
\begin{equation}
f'(\vec{\theta})=P(1)-P(0)=-|\langle \psi(\vec{\theta})|U|\psi(\vec{\theta})\rangle|^2
\end{equation}
or 
\begin{equation}
f''(\vec{\theta})=\frac{1}{f(\vec{\theta})}=\frac{1}{P(0)-P(1)}.
\end{equation}
With these functions that can be computed from measurements on the circuit, we can use a classical optimization methods to update the parameters $\vec{\theta}$ until the parametrized circuit giving $\mathbf{P}(\vec{\theta})$ produces a good approximation to an eigenstate of the target unitary $U$.

\subsection{Solving the generalized eigenvalue problem }
\label{generalized}
\begin{figure}[h]
\mbox{
\Qcircuit @C=1em @R=.7em {
\lstick{\ket{0}} &   \qw         &  \multigate{3}{\mathbf{P}( \vec{\theta}\,)}  	& \qw  		&  \multigate{3}{\,U\,}	& \qw 		& \ctrl{5}	& \qw  		& \qw 		& \dots & & \qw  & \qw &\gate{H}  & \meter &\rstick{O_1}\\
\lstick{\ket{0}} &   \qw         & \ghost{\mathbf{P}( \vec{\theta}\,)} 		& \qw 		&  \ghost{\,U\,}  	& \qw 		& \qw  		& \ctrl{5}	& \qw 		& \dots & & \qw  &\qw  &\gate{H} & \meter &\rstick{O_2}\\
\push{}          & \push{\vdots} &  \push{}  				& \push{\vdots} & \push{} 		& \push{} 	& \push{}	& \push{}	& \push{}       & \push{\ddots} 	& \push{} & \push{}& \push{} & \push{} &  \push{\vdots}\\
\lstick{\ket{0}} &   \qw         & \ghost{\mathbf{P}( \vec{\theta}\,)}   		& \qw 		&   \ghost{\,U\,}  	& \qw 		&  \qw  	&  \qw 		&\qw            & \dots &  & \ctrl{5}  &\qw  &\gate{H} & \meter &\rstick{O_n}\\
\push{}          & \push{}       & \push{} 				&  \push{} 	& \push{} 		& \\
\lstick{\ket{0}} & \qw           &  \multigate{3}{\mathbf{P}( \vec{\theta}\,)}  	& \qw 		& \multigate{3}{\,V\,}				&\qw  		& \targ  	& \qw  		& \qw 		& \dots & & \qw  & \qw  &\gate{H} & \meter &\rstick{O_1'}\\
\lstick{\ket{0}} & \qw           & \ghost{\mathbf{P}( \vec{\theta}\,)}  		& \qw 		& \ghost{\,V\,}		& \qw  		& \qw  		& \targ 	&\qw 		& \dots & & \qw  & \qw   &\gate{H} & \meter &\rstick{O_2'}\\
\push{}          & \push{\vdots} &  \push{} 				& \push{\vdots} & \push{}	        & \push{}		& \push{} 	& \push{}	& \push{} & \push{\ddots} & \push{} & \push{}& \push{} & \push{} &  \push{\vdots}\\
\lstick{\ket{0}} & \qw           & \ghost{\mathbf{P}( \vec{\theta}\,)}  		& \qw 		&  \ghost{\,V\,} 	& \qw 		& \qw 		&  \qw 	 	& \qw           & \dots & &  \targ& \qw   &\gate{H}& \meter &\rstick{O_n'}\\
}
}
\caption{Variational quantum circuit for the generalized eigenvector problem $U\ket{e_i}=\lambda V\ket{e_i}$.}\label{Generalized}
\end{figure}
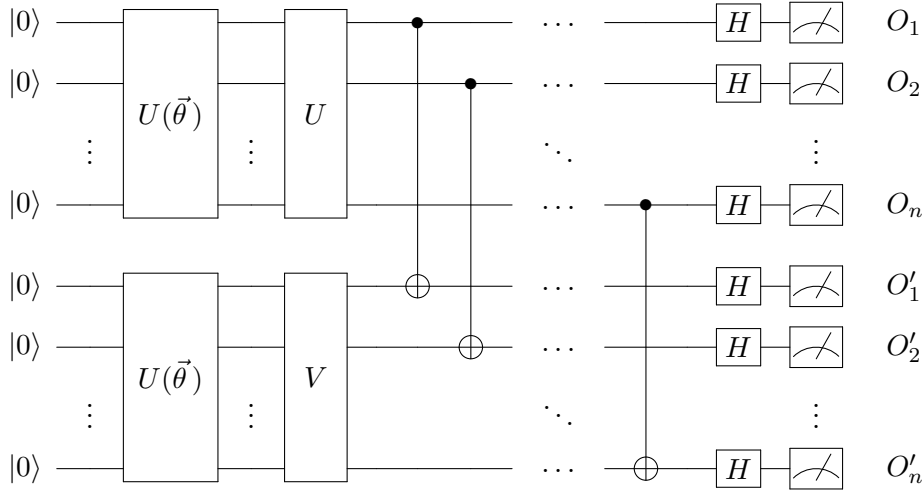

The circuit can be modified to find generalized eigenvectors in what is usually called the generalized eigenvalue problem of finding $\ket{\psi}$ and $\lambda$ such that
\begin{equation}
U\ket{e_i}=\lambda V\ket{e_i}
\end{equation}
for two given matrices $U$ and $V$ (see Figure \ref{Generalized}). The hybrid algorithm is exactly the same as before, now maximizing the objective function $f(\vec{\theta})=|\langle \psi(\vec{\theta})|U^\dag V|\psi(\vec{\theta})\rangle|^2$ until $f(\vec{\theta})=|\bra{e_i}U^\dag V\ket{e_i}|^2=|\bra{e_i}\lambda^*\ket{e_i}|^2=1$.

\subsection{Generalization for normal matrices}
\label{normal}
The variational eigenvector finder can be modified to work on normal matrices for both the usual and the generalized eigenvalue problems. Consider a normal matrix $N$ such that $NN^\dag=N^\dag N$. For an $M\times M$ normal matrix the spectral theorem guarantees we have a decomposition:
\begin{equation}
\label{SpecDec}
N=\sum_{i=1}^M \lambda_i \ket{e_i}\bra{e_i}
\end{equation}
for $M$ orthogonal eigenvectors. This spectral decomposition might not be unique if there are degenerate eigenvalues. 

For any square $M\times M$ complex matrix $A$, the matrix $H_A=A+A^\dag$ is Hermitian ($H_A=H_A^\dag$). If $A$ admits a spectral decomposition like in Eq. (\ref{SpecDec}), this Hermitian matrix shares the eigenvectors of $A$ with
\begin{equation}
H_A=A+A^\dag=\sum_{i=1}^M \lambda_i \ket{e_i}\bra{e_i}+\sum_{i=1}^M \lambda_i^* \ket{e_i}\bra{e_i}=\sum_{i=1}^M (\lambda_i+\lambda_i^*) \ket{e_i}\bra{e_i}=\sum_{i=1}^{M} 2\operatorname{Re}(\lambda_i) \ket{e_i}\bra{e_i}.
\end{equation}

If we consider the unitary $U_A=e^{iH_A}$, it will have a decomposition
\begin{equation}
U_A=\sum_{i=1}^{M} e^{i2\operatorname{Re}(\lambda_i)} \ket{e_i}\bra{e_i}.
\end{equation}

The matrices $A$, $H_A$ and $U_A$ have the same eigenvectors with eigenvector $\ket{e_i}$ associated to eigenvalues $\lambda_i$, $2\operatorname{Re}(\lambda_i)$ and $e^{i2\operatorname{Re}(\lambda_i)}$, respectively.

If we can construct the unitary evolution $U_A$, we can obtain the eigenvectors of any $A$ admitting a spectral decomposition, including all normal matrices. Finding an efficient quantum gate sequence that realizes the exponential $e^{iH_A}$ might be difficult. This is a well-studied problem that appears in Hamiltonian simulation \cite{Llo96,BCC15} and in the HHL quantum algorithm for linear systems of equations \cite{HHL09}. All the tricks used there can be recycled in this variational method, including efficient approximation for sparse matrices \cite{BCC14} or designing evolutions directly for the particular quantum computer architecture in which the variational circuit is run \cite{CBC21}.

\section{Principal Component Analysis of unknown mixed states}
\label{PCA}
The proposed variational method can be also extended to statistical mixtures, which can appear, among other situations, after a pure state is subject to decoherence. Any mixed state can be described by a density matrix 
\begin{equation}
\rho=\sum_{j}^{m} p_j \ket{\psi_j}\bra{\psi_j},
\end{equation}
with a variable number of terms $m$ where the $p_j$ are the probabilities associated to finding a pure state $\ket{\psi_j}$ and sum to one. This decomposition is not unique and the states are not necessarily orthogonal. The density matrix $\rho$ is still Hermitian and it will have a spectral decomposition
\begin{equation}
\label{spectraldec}
\rho=\sum_{j=1}^{N} \lambda_j \ket{\psi_i}\bra{\psi_i},
\end{equation}
where $N$ is the dimension of the state space and the $\ket{\psi_i}$ are now orthogonal states. For non-degenerate eigenvalues, this decomposition is unique.

One important problem for mixed states is Quantum Principal Component Analysis (QPCA): determining the leading terms in the spectral decomposition. The eigenvalues of the Hermitian $\rho$ are real and can be ordered from the largest to the smallest as $\lambda_1\geq \lambda_2 \geq \ldots \geq \lambda_N$. The principal component is the eigenstate associated to the largest eigenvalue which, in certain scenarios, gives a good approximation to the full state. This mirrors classical Principal Component Analysis methods \cite{Pea01}, which have applications in multiple fields, like in machine learning, where it is generally used as a way to reduce the dimensionality of the data \cite{JC16,Jol02}.

The general QPCA problem is thought to be hard for classical computers. Lloyd, Mohseni and Rebentrost showed that there is an efficient quantum algorithm for QPCA as long as we can prepare multiple copies of a state with a density matrix $AA^\dag$ for any given matrix $A$ \cite{LMR14}. The proposed quantum method was later dequantized by Tang who showed that, if we assume there is an equivalent black box giving this state preparation, there are classical methods which are only quadratically worse than the quantum proposal (as opposed to the exponential quantum advantage in the case of low rank matrices) \cite{Tan21}.

The variational quantum algorithm proposed in the previous sections can also be used to perform principal component analysis with the same caveats regarding state preparation. We assume we are given multiple copies of a mixed state with density matrix $\rho$. While for general tasks it might be hard mapping a classical matrix to a state, the method can still be useful, for instance, to approximate the input of a noisy channel within certain precision when the principal component does indeed give a good estimation of the original input \cite{Koc21}. When compared to the QPCA algorithm of \cite{LMR14}, we just need the raw input states instead of using them to build the evolution $e^{i\rho}$, which makes the variational algorithm more suitable for current noisy quantum computers. However, we are subject to the usual problems of variational methods: they are heuristic and we have no guarantee of finding a solution, among other challenges.  

\begin{figure}[h]
\mbox{
\Qcircuit @C=1em @R=.7em {
& \lstick{\ket{0}} & \gate{H} & \qw &\qw &\qw&\ctrl{1} &\gate{H} & \meter \\
& \lstick{\ket{0}^{\otimes n}} & \qw{/}  & \gate{\mathbf{P}(\,\vec{\theta}\,)}& \ustick{\scriptstyle{\sigma(\vec{\theta})}} \qw & \qw{/}  & \multigate{1}{SWAP}&\qw{/}&\qw \\
& \lstick{\rho} &  \qw & \qw{/} & \qw & \qw&\ghost{SWAP}& \qw{/}&\qw
}
}
\caption{Quantum circuit for variational quantum principal component analysis.}\label{SWAPtestcircuitPCA}
\end{figure}
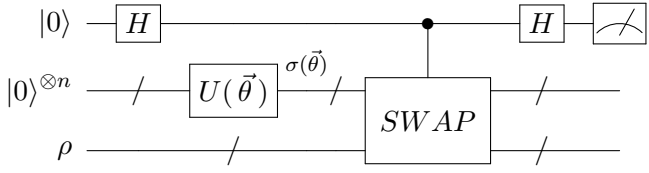

We will use the circuit in Figure \ref{SWAPtestcircuitPCA}, which also admits a gate efficient realization with a destructive SWAP test like the circuits in Figures \ref{VarU} and \ref{Generalized}. For an unknown input mixed state $\rho$, we perform a series of SWAP tests comparing the input to the pure state $\sigma(\vec{\theta})=|\psi(\vec{\theta})\rangle\langle\psi(\vec{\theta})|$ generated from a fixed input and a parametrized quantum circuit controlled by the classical parameters $\vec{\theta}$. In the optimization phase, we will tweak these classical parameters until the \emph{ansatz} $\sigma(\vec{\theta})$ maximizes the value of the measured observable.

From Equations (\ref{prob0mix}) and (\ref{spectraldec}), we see the expected value we approximate is 
\begin{equation}
\label{mixSWAP}
\tr\left( \rho \sigma(\vec{\theta})  \right)=\tr\left(\sum_{i=1}^N \lambda_i \langle\psi(\vec{\theta})\ket{\psi_i}\bra{\psi_i}\psi(\vec{\theta})\rangle \right)=\sum_{i} \lambda_i |\langle\psi(\vec{\theta})\ket{\psi_i}|^2.
\end{equation}
The global maximum is achieved when the ansatz is the principal component (the eigenstate $\ket{\psi_1}$ associated to the largest eigenvalue $\lambda_1$). In that case the expected value becomes $\lambda_1$. For any other trial state, with at least some part of the state orthogonal to $\ket{\psi_1}$ and $\lambda_1 \geq \lambda_2 \geq \lambda_3\geq \cdots \geq \lambda_N$:
\begin{equation}
\tr\left( \rho \sigma(\vec{\theta})  \right)=\sum_{i} \lambda_i |\langle\psi(\vec{\theta})\ket{\psi_i}|^2\leq \lambda_1 \sum_{i} |\langle\psi(\vec{\theta})\ket{\psi_i}|^2 = \lambda_1. 
\end{equation}
In general, the final state found at the end of the algorithm will depend on the optimization method and the input state. If we start close to a principal component the objective function will find a local maximum with a trial state close to the desired state. However, the search algorithm can also converge to local maxima that do not correspond to a principal component. The global maximum will be more prominent for mixed states with a strong principal component.

There are two important differences with respect to eigenvector estimation for unitary and normal matrices. In those cases, we could know for sure the algorithm had converged to a true eigenvector by checking the expected value, which should converge to 1. Here we cannot be sure we have found the principal component and not converged to some local maximum (likely a second or third component, or a superposition of the first components). However, we get additional information if we did. In QPCA, it is useful finding both the state and the eigenvalue, which can be learnt from the expected value. The value will tell us how important the principal component is with respect to other terms and how well it can be used as a compact description of the whole state. Larger values correspond to principal components that represent the whole mixed state better. 

Using multiobjective optimization, we can also search for successive components. When the algorithm converges after $k$ iterations, we obtain a set of parameters $\vec{\theta}_k^1$ that give the estimated principal component $|\tilde{\psi_1}\rangle=|\psi(\vec{\theta}_k^1)\rangle$. We can now alternate two series of rounds for each iteration $j$ in order to evaluate the parameters $\vec{\theta}_j^2$ defining our trial state for the second component. First we can perform the usual comparison of the \emph{ansatz} with the input state $\rho$ to estimate $\tr(\rho\sigma(\vec{\theta}_j^2))$. Then, we generate as input states $\sigma(\vec{\theta}_k^1)$ and $\sigma(\vec{\theta}_j^2)$ and use a SWAP test to estimate the overlap $\tr(\sigma(\vec{\theta}_k^1)\sigma(\vec{\theta}_j^2))=|\langle\psi(\vec{\theta}_k^1)|\psi(\vec{\theta}_j^2)\rangle|^2$. We want two orthogonal states such that $\tr(\sigma(\vec{\theta}_k^1)\sigma(\vec{\theta}_j^2))=0$. Using these two estimated values, the classical parameters must be optimized according to a new objective function that the classical algorithm will use to propose the vector $\vec{\theta}_{j+1}^2$ for the next iteration. Assuming we minimize $f(\vec{\theta})$ during optimization, we can, for instance, define an objective function
\begin{equation}
\label{multiobjective}
f(\vec{\theta})=1/\tr( \rho \sigma(\vec{\theta})  )+C|\langle\psi(\vec{\theta})|\tilde{\psi_1}\rangle|^2
\end{equation}
with a weight constant $C$ that penalizes the previously found principal component and directs the search towards the next maximum eigenvalue in the decomposition of $\rho$. This procedure can be repeated as many times as needed with additional terms to search for the third, fourth and successive components, much like molecular search for excited energy levels above the ground state in molecules \cite{HWB19}. 

\section{Optical implementation}
\label{optical}
The variational algorithms for eigenvectors and quantum principal component analysis we have seen are not restricted to near-term universal quantum computers. They can also be carried out in linear optical setups with certain restrictions. 

We consider passive linear optical systems with input states that are a superposition of terms $\ket{n_1n_2\cdots n_m}$ where we have $n=\sum_{i=1}^{m}n_i$ photons distributed into $m$ orthogonal modes (like different paths or orthogonal polarizations). A linear optics system acting on $m$ modes can be described classically by its scattering matrix $S$ which is an $m\times m$ unitary matrix \cite{Hau95,Poz04}. The corresponding quantum evolution acting on $n$ photons can be computed from $S$ as $U=\varphi_{m,n}(S)$ using different known methods \cite{Cai53,SGL04,Sch04,AA11}. The unitary $U$ has a size $M\times M$, with $M={{n+m-1}\choose{n}}$. For one photon the quantum evolution is exactly the scattering matrix $S=U$ \cite{CAK98}.

There are multiple constructive methods to map any desired unitary $S$ into a physical setup using only linear optical devices such as beamsplitters and phase shifters \cite{RZB94,BA14,Saw16,CHM16,GMS18,BW21}. These implementations have a circuit size that grows quadratically with the number of modes in the worst case and can generate any desired unitary $S$. There exist many successful experimental realizations of configurable optical circuits with integrated optics \cite{CHS15,MCS18,BPC20,EPS20,ABB21,TMS21,HPG22} that offer a linear optical system that can be controlled electronically. These hybrid systems have configurable phase shifters and beamsplitters and give quantum evolutions that depend on a few classical parameters. They can be designed to provide any available scattering matrix for the size input, but there are also simpler configurable circuits that can produce good enough \emph{ans\"atze} in variational quantum algorithms \cite{PMS14}.

These systems give an efficient way to realize the first half of the proposed quantum circuits which has the parametrized unitaries that generate the trial states for eigenvector estimation and principal component analysis. For the second half which performs the state comparison, we need to restrict our setups. 

The SWAP test admits a simple experimental realization for two photons. We start from two separate single photons in the same large Hilbert space. We just need to direct the two photons encoding the states to the two inputs of a balanced beamsplitter. At the output of the beamsplitter we place two binary photodetectors (that click for one or more photons and do nothing for the vacuum). In such a setup, we call a coincidence to the simultaneous detection of one photon at each output port of the beamsplitter. For two input single photon states described by density matrices $\rho$ and $\sigma$, the probability of finding a coincidence at the output is \cite{SvE11}:
\begin{equation}
P_C=\frac{1-\tr(\rho\sigma)}{2},
\end{equation}
which is exactly the probability of failing a SWAP test. That way, optical measurements at a beamsplitter can be used for state comparison \cite{GC13}.

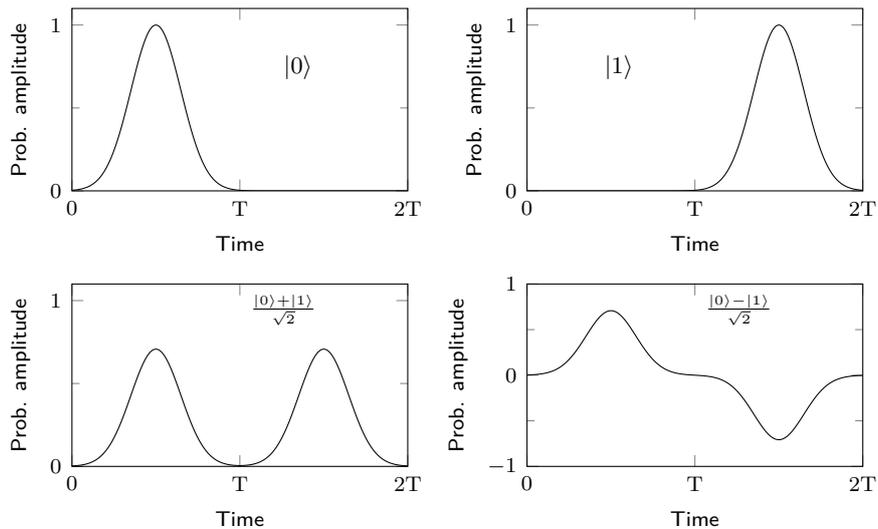
\begin{figure}
\centering
\pgfplotsset{every tick label/.append style={font=\scriptsize}}
\begin{tikzpicture}[font=\sffamily,/pgfplots/.cd,width=6cm,height=4cm]
\matrix{
 \begin{axis}[xmin=-4,xmax=4,ymin=0,ymax=1.1,
    xtick={-4,0,4},
    xticklabels={0,T,2T},
    ytick={0,1},
    minor y tick num=1,
    xlabel={\scriptsize{Time}},ylabel=\scriptsize{Prob. amplitude},title={\footnotesize $\ket{0}$}, every axis title/.style={below right,at={(0.6,0.8)}}]
  \addplot[no markers,smooth,samples=201,domain=-4:4] 
    {exp(-(x+2)*(x+2)*sqrt(2))};
 \end{axis}&[1mm]
 \begin{axis}[xmin=-4,xmax=4,ymin=0,ymax=1.1,
    xtick={-4,0,4},
    xticklabels={0,T,2T},
    ytick={0,1},
    minor y tick num=1,
    xlabel={\scriptsize{Time}},ylabel=\scriptsize{Prob. amplitude},title={\footnotesize $\ket{1}$}, every axis title/.style={below right,at={(0.2,0.8)}}]
  \addplot[no markers,smooth,samples=201,domain=-4:4] 
    {exp(-(x-2)*(x-2)*sqrt(2))};
 \end{axis}\\[1mm]
 \begin{axis}[xmin=-4,xmax=4,ymin=0,ymax=1.1,
    xtick={-4,0,4},
    xticklabels={0,T,2T},
    ytick={0,1},
    minor y tick num=1,
    xlabel={\scriptsize{Time}},ylabel=\scriptsize{Prob. amplitude},title={\tiny $\frac{\ket{0}+\ket{1}}{\sqrt{2}}$}, every axis title/.style={below right,at={(0.5,1)}}]
  \addplot[no markers,smooth,samples=201,domain=-4:4] 
    {1/sqrt(2)*exp(-(x+2)*(x+2)*sqrt(2))+1/sqrt(2)*exp(-(x-2)*(x-2)*sqrt(2))};
 \end{axis}&[1mm]
 \begin{axis}[xmin=-4,xmax=4,ymin=-1,ymax=1,
    xtick={-4,0,4},
    xticklabels={0,T,2T},
    ytick={-1,0,1},
    minor y tick num=1,
    xlabel={\scriptsize{Time}},ylabel=\scriptsize{Prob. amplitude},title={\tiny $\frac{\ket{0}-\ket{1}}{\sqrt{2}}$}, every axis title/.style={below right,at={(0.5,1)}}]
  \addplot[no markers,smooth,samples=201,domain=-4:4]   
    {1/sqrt(2)*exp(-(x+2)*(x+2)*sqrt(2))-1/sqrt(2)*exp(-(x-2)*(x-2)*sqrt(2))};
 \end{axis}\\
};
\end{tikzpicture}
\caption{Time bin encoding for optical qubits. The time location of an optical pulse can serve as a basis for a qubit. With different amplitude combinations and relative phases between the optical pulses, we can produce any desired qubit.}
\label{opticalqubit}
\end{figure}

Notice we need to encode the whole state from a large Hilbert space into a single photon. There are many alternatives, like taking advantage of the high dimensionality provided by orbital angular momentum \cite{ABS92,ABP03,WPS21}. A good option compatible with current optical network technology is using a time-bin encoding for the orthogonal modes. Figure \ref{opticalqubit} shows different states of a qubit in time-bin encoding. The photon wavefunction can be confined to two different time bins representing the $\0$ and $\1$ qubits. If we use a common phase reference that is assigned to the pulses in the $\ket{0}$ time bin, we can use amplitude modulators to distribute the probability amplitude in the two time bins and phase modulators to introduce a relative phase with respect to the reference. The result is an optical qubit in any state of choice. This approach has been used, for instance, in quantum key distribution protocols over optical fiber \cite{MHH97,WLP21}.

In the proposed variational algorithms, we need $d$-dimensional systems, or qudits, which can be generated by dividing the photon into $d$ time bins. With the current technology, we can expect a reasonable coherence time of microseconds and amplitude and phase modulators in the GHz range, which could produce systems with 1024 bins, equivalent to 10 qubits. Existing modulators, beamsplitters and detectors could give results comparable to what can be done in a quantum computer with 20 qubits. Each of the two photons could also be distributed into multiple separate spatial modes, possibly with added time-bin encoding. The measurement would need a balanced beamsplitter for each pair of modes corresponding to the same position for each photon and one detector at each of their outputs.

In all the variational algorithms proposed, both for finding eigenvectors and for quantum principal component analysis, one of the subsystems would be one photon directed to a parametrized linear optics multiport. The second system can either come from a general optical quantum channel, which could produce a mixed state we want to characterize, or a known optical transformation if we want to map a known matrix $S$ into the optical system in order to find a compact description of its eigenvectors, as, in this case with a single photon, $U=S$. 

While the scalability of this approach is limited, it can be an interesting alternative to full quantum computers in the near term, particularly in the cases where we need a low noise implementation. Photons have long coherence times, relatively low noise and do not need advanced cooling systems unlike many implementations of quantum computers, especially those based on superconducting qubits. Optical systems have their own problems like losses, synchronization or producing good approximations to single photon states, but, for most of these problems, there are acceptable technical solutions that have been used in quantum key distribution systems \cite{SBC09,XMZ20}.

\section{Comparison to other variational quantum algorithms for eigenvector determination}
\label{comparison}
There exist various proposals for variational quantum algorithms that learn an eigenstate of a Hamiltonian, a density matrix or a general matrix, each focusing on different objectives. The main motivation of the algorithms put forward in this paper is to reduce as much as possible the depth of final quantum circuit.

Noisy intermediate scale quantum computers are the natural hardware to run the quantum part of hybrid variational quantum algorithms. However, variational methods are severely limited by the noise in each quantum gate. 

A first strong limitation is that there is a bound to the maximum depth for which variational algorithms give genuine quantum advantage \cite{SG21}. Above a certain depth threshold, which depends on the noise level, classical Monte Carlo techniques offer an efficient simulation of the quantum variational method. Even if the results are correct despite the noise, the problems that can be solved in this case can also be solved, without all the specialized hardware, using a classical computer.

A second important limitation comes from the classical optimization phase. Noise introduces barren plateaus in the optimization landscape \cite{WFC21}: the gradient vanishes exponentially with the depth of the parametrized circuit generating the ansatz. This makes classical optimization unfeasible. The search for a global minimum of the cost function will get stuck at some sub-optimal parameters.

We can compare the proposed variational methods to previous algorithms in terms of the depth of the circuits.

We have approximated the overlap $\bra{\psi(\vec{\theta})}U\ket{\psi(\vec{\theta})}$ comparing two states $U\ket{\psi(\vec{\theta})}=U\mathbf{P}(\vec{\theta})\ket{0}^{\otimes n}$ and $\ket{\psi(\vec{\theta})}=\mathbf{P}(\vec{\theta})\ket{0}^{\otimes n}$ using the SWAP test. However, we can just take an initial all-zeroes state and make it evolve under $\mathbf{P}^\dag(\vec{\theta})U\mathbf{P}(\vec{\theta})$ \cite{KLP19}. If we project the resulting state to the $\ket{0}^{\otimes n}$ output, the statistics for finding all the output qubits in $\ket{0}$ are $\hspace{-2.5ex}{\phantom{\bra{0}}\!}^{n\otimes }\!\bra{0}\mathbf{P}^\dag(\vec{\theta})U\mathbf{P}(\vec{\theta})\ket{0}^{\otimes n}$ which is exactly $\bra{\psi(\vec{\theta})}U\ket{\psi(\vec{\theta})}$.

This approach has advantages and disadvantages with respect to using the SWAP test. The biggest advantage is that is only uses $n$ qubits for $2^n\times 2^n$ unitaries instead of $2n$ qubits and does not require any additional gates. In the proposed destructive SWAP test the overhead of the SWAP test reduces to two gates per qubit pair, which is not so important. The qubit advantage is not trivial, though, particularly for intermediate scale applications where qubits are scarce. However, the additional circuit depth due to including $\mathbf{P}(\vec{\theta})$ can be important, as the performance of variational algorithms can degrade exponentially with the total circuit depth \cite{WFC21} and deep noisy circuits can loose quantum advantage \cite{SG21}.

There are also other variational methods that assume a known Hamiltonian that can be efficiently written in the Pauli basis \cite{KMvB19,KUY20}. For eigenstates, the cost function
\begin{equation}
C(\vec{\theta})=\bra{\psi(\vec{\theta})}H^2\ket{\psi(\vec{\theta})}-\bra{\psi(\vec{\theta})}H\ket{\psi(\vec{\theta})}^2
\end{equation}
is equal to zero and can be used to tune the parameters in $\mathbf{P}(\vec{\theta})$. This only requires $n$ qubits for a Hamiltonian corresponding to a $2^n\times 2^n$ matrix. The additional cost of measuring the corresponding Pauli operators is small (at most a change of basis gate on each qubit). This can be more efficient than the SWAP test eigenvector finder, but it only works for known Hamiltonians which can be efficiently decomposed into Pauli operators (in some encoding). 

Notice that the SWAP test algorithms proposed in this paper would also work on arbitary unitaries $U$, even if we just have a description in terms of the quantum circuit giving the evolution. We don't need to compute a closed form for the matrix $U$ or find the corresponding Hamiltonians.

The application of the SWAP test for variational quantum principal analysis works on unknown input density matrices $\rho$ and has the simplest circuit from all we have seen in this paper. Two notable variational algorithms are given in \cite{LTO19} and \cite{CSA22}, both of which follow the same philosophy of the method in Section \ref{PCA}: after the procedure we have a set of classical parameters that allow the reconstruction of the desired principal component with a small error. 

The algorithm for variational quantum state diagonalization in \cite{LTO19} takes two copies of the unknown state per each quantum iteration. Both copies go through the same parametrized diagonalization circuit $\mathbf{P}(\vec{\theta})$ and they are compared using variations of the destructive SWAP test or inner product measurements. The cost function is chosen so that, after successful optimization, $\mathbf{P}^\dag(\vec{\theta})\rho\mathbf{P}(\vec{\theta})$ is diagonal. Later, this diagonalizing operation can be used to recover the eigenvalues and eigenvectors of $\rho$. The complexity is essentially the same as the circuit in Figure \ref{SWAPtestcircuitPCA}. The circuits have the same depth, but the method needs two copies of the parametrized quantum circuit and the input state instead of the one copy in Figure \ref{SWAPtestcircuitPCA}.

The alternative variational quantum state eigensolver in \cite{CSA22} only requires one copy of $\rho$ and the parametrized circuit and, instead of the SWAP test approach, it only uses $n$ qubits (for states in a Hilbert space of dimension $2^n$). This algorithm produces the same result as the method in \cite{LTO19} using a cost function $C(\vec{\theta})=\tr(\mathbf{P}^\dag(\vec{\theta})\rho\mathbf{P}(\vec{\theta}) H)$ for a given Hamiltonian $H$ which can be adapted during training. This method is more versatile and compact at the cost of adding a few gates to include the Hamiltonian in the measurement. 

In general, these two algorithms are more flexible than the method in Section \ref{PCA} and they give a full solution to principal component analysis with a slightly higher gate cost. If the full spectrum is needed they can be more interesting, but they both require more sophisticated processing and interpretation.

\section{Strong and weak points, applications and outlook}
\label{discussion}
Classically, finding the eigenvalues of a matrix is an efficient task. For unitary matrices the QR decomposition \cite{Fra61,Fra62,Kub62} gives a robust algorithm with efficient and stable numerical software implementations in reference suites like the LAPACK library \cite{ABB99}. For an $N\times N$ matrix, the number of operations grows as $O(N^3)$, with alternative algorithms with complexities between quadratic and cubic depending on special cases \cite{Dem97,Ste01,Bjo15}. This includes bisection methods that can search only for one or a limited subset of eigenvectors instead of the whole spectrum at a smaller cost in terms of the number of operations. 

However, in a quantum setting, the size of the state space grows exponentially in the number of qubits. For $n$ qubits this means that any algorithm that needs to explicitly write down the $N=2^n$ complex entries in a general eigenvector will suffer from this growth. The variational quantum algorithms introduced in this paper return a compressed version of the state so that we can produce the desired eigenstate on demand. The classical parameters in $\vec{\theta}$ are a short list that contains all the information we need to recreate the desired eigenvector using the parametrized circuit we chose for the search. This native encoding into a quantum state allows for an efficient preparation for future uses. 

The algorithm for variational quantum principal component analysis can also be useful for channel characterization, for instance in the analysis of mixed states coming from a quantum optical channel with decoherence or to study the decay of a quantum state with time in a noisy quantum computer.

The main potential application for these methods is finding eigenvectors as a stepping stone for more complex routines. Take for instance the Quantum Phase Estimation algorithm of Kitaev \cite{Kit95} where a quantum computer can give efficient approximations to an eigenvalue of a given unitary provided one of the inputs is the corresponding eigenvector. The variational quantum eigenvector finder can be used in conjunction with Quantum Phase Estimation to obtain the full spectrum of any unitary $U$, including all the eigenvalues.

The variational eigenvector finder also combines well with Abram and Lloyd's algorithm that gives an efficient approximation to the eigenvectors of a Hamiltonian with arbitrary precision as long as it has a ``good enough'' initial guess state \cite{AL99}. If we can produce a state $\ket{\tilde{e_i}}$ that approximates the desired exact eigenstate $\ket{e_i}$ and has a non-negligible overlap $|\langle \tilde{e_i}| e_i \rangle|^2$ with it, the eigenstate can be refined and taken as close to the exact value as desired. This fits particularly well with the output of the variational eigenvector finder. Even if the optimization does not converge to an exact eigenvector, we can determine from $P(0)-P(1)$ (the $Z$ observable from the SWAP test) whether the trial state is close or not to an eigenstate. Larger expected values correspond to a greater overlap. With the correction from Abrams-Lloyd algorithm, even poor \emph{ans\"atze} can give a clean eigenstate. The

While these are interesting applications, both require long circuits and use the Quantum Fourier Transform which, for large systems, is still impractical due to accumulated noise. A more realistic application in the short term would be optimizing algorithms for noisy intermediate scale quantum computers.

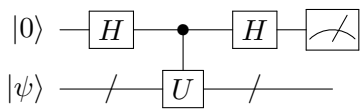
\begin{figure}[h]
\mbox{
\Qcircuit @C=1em @R=.7em {
& \lstick{\ket{0}} & \gate{H} & \ctrl{1} &\gate{H} & \meter \\
& \lstick{\ket{\psi}} & {/}\qw & \gate{U}&\qw{/}&\qw\\
}
}
\caption{Quantum circuit realizing the Hadamard test.}\label{Hadamardtestcircuit}
\end{figure}

Take for instance the Hadamard test of Figure \ref{Hadamardtestcircuit}, which appears in many proposals for quantum machine learning \cite{BLS19,CWV20} and in the quantum algorithm for Jones polynomials \cite{AJL06,PMR09}. This circuit is a simpler version of the Quantum Phase Estimation algorithm that still can give results that no classical system can.  

A key element in that circuit is the controlled unitary $cU$ which applies a unitary $U$ to the target when the control qubit is $\1$ and is the identity for a control qubit $\0$. If we are given $U$ as a black box, this task is impossible in the usual quantum circuit model without describing the complete $cU$ operator and decomposing it \cite{AFC14}. The total number of gates in those decompositions can be large, introducing too much noise, and there have been various attempts to give concise descriptions for particular cases in order to reduce the depth of these circuits \cite{LLS08,MF19,AA21,WRZ21}. In particular, if we can produce an eigenstate of the evolution there are restricted efficient circuits for $cU$ \cite{Kit95}. The variational quantum eigenvector finder can be part of general methods to compute a Hadamard test with a smaller total number of gates.

A preliminary simulation of the proposed quantum variational eigenvector finder \cite{Gar23} shows that, for moderate sizes of $U$, the algorithm converges to valid eigenvectors both for random input unitaries and for defective matrices with high degeneracy like the Quantum Fourier Transform. As the size of the problem grows, the found state has a smaller overlap with a true eigenstate. This is a common problem in variational quantum algorithms. For an ideal algorithm, we can just use more complex parametrized circuits to cover a larger part of the state space up to a point. However, in noisy execution, this approach can be counterproductive. In the simulations, even for the smallest problems, noise introduces an appreciable deviation from true eigenstates. 

On this account, the proposed algorithms share the drawbacks of most hybrid variational methods. First of all we need a good \emph{ansatz}. There are different proposals for parametrized circuits that could be used. In any variational method there is a tension between expressivity and depth. Expressivity describes how much of the whole state space can be reached from outputs of the parametrized circuit. Depth is given by the number of consecutive elementary gates we need to build the circuit. We would like to have expressive enough circuits that can generate an ansatz close to our target state (the exact eigenvector we are searching for). However, highly expressive parametrized circuits tend to have larger depths and the noise can build up to levels that make the output unusable. Apart from that, in order to sample more of the space they need more classical parameters. Optimizing a large set of parameters is less efficient and the classical part of the algorithm can converge to suboptimal solutions. Among other challenges, if we can represent very small changes, there can appear barren plateaus (the optimization algorithm gets stuck in a region of the state space with small changes in the objective function but far from the real minimum) \cite{HSC22}. While this can limit the usefulness of the method as the state space grows, there are methods to optimize the parametric circuits so that the number of parameters is small but we are still able produce states close to the target \cite{DTY22,FHJ22}. The search for better parametrized circuits and optimization methods is a vibrant area of research and most of the results for other variational quantum algorithms are likely to be useful in the eigenvector algorithms. 

One advantage with respect to other problems, like searching for the ground state in molecular simulation, is that we do have a confirmation whether the algorithm has succeeded or not. We might not converge to a true solution, but, when we do, the expected value will be one.  

In general, hybrid algorithms are an interesting solution to middle sized problems, but have challenge scaling to really large state spaces. The classical optimization phase becomes more involved with the number of parameters and we might not converge to the optimal solution. Even in this case, if scalable full quantum computers become available, the initial approximation the variational eigenvector finder gives can be used to kickstart the Abrams-Lloyd algorithm \cite{AL99}.

Despite these limitations, the presented variational eigenvector finder and quantum principal component analysis algorithms can become a useful addition to the toolbox of quantum computation in the near future and they can be incorporated into practical, more complex hybrid quantum-classical algorithms.

\section*{Acknowledgments}
This work has been funded by the Spanish Ministerio de Ciencia e Innovaci\'on (MCIN), project PID2020-119418GB-I00, and by the European Union NextGeneration UE/MICIU/Plan de Recuperaci\'on, Transformaci\'on y Resiliencia/Junta de Castilla y Le\'on.


\begin{thebibliography}{10}

\bibitem{Sho97}
P.W. Shor.
\newblock ``Polynomial-time algorithms for prime factorization and discrete
  logarithms on a quantum computer''.
\newblock \href{https://dx.doi.org/10.1137/S0097539795293172}{\mbox{SIAM
  Journal on Computing} {\bf 26}, 1484}~(1997).

\bibitem{Mon16}
Ashley Montanaro.
\newblock ``Quantum algorithms: an overview''.
\newblock \href{https://dx.doi.org/10.1038/npjqi.2015.23}{npj Quantum
  Information {\bf 2}, 1--8}~(2016).

\bibitem{Pre18}
John Preskill.
\newblock ``Quantum {C}omputing in the {NISQ} era and beyond''.
\newblock \href{https://dx.doi.org/10.22331/q-2018-08-06-79}{{Quantum} {\bf 2},
  79}~(2018).

\bibitem{BCK22}
Kishor Bharti, Alba Cervera-Lierta, Thi~Ha Kyaw, Tobias Haug, Sumner Alperin-Lea,
 Abhinav Anand, Matthias Degroote, Hermanni Heimonen, Jakob~S. Kottmann, Tim Menke,
 Wai-Keong Mok, Sukin Sim, Leong-Chuan Kwek, and Al{\'a}n Aspuru-Guzik.
\newblock ``Noisy intermediate-scale quantum algorithms''.
\newblock \href{https://dx.doi.org/10.1103/RevModPhys.94.015004}{Reviews of
  Modern Physics {\bf 94}, 015004}~(2022).


\bibitem{MRB16}
Jarrod~R McClean, Jonathan Romero, Ryan Babbush, and Alán Aspuru-Guzik.
\newblock ``The theory of variational hybrid quantum-classical algorithms''.
\newblock \href{https://dx.doi.org/10.1088/1367-2630/18/2/023023}{New Journal
  of Physics {\bf 18}, 023023}~(2016).

\bibitem{ECB21}
Suguru Endo, Zhenyu Cai, Simon~C. Benjamin, and Xiao Yuan.
\newblock ``Hybrid quantum-classical algorithms and quantum error mitigation''.
\newblock \href{https://dx.doi.org/10.7566/JPSJ.90.032001}{Journal of the
  Physical Society of Japan {\bf 90}, 032001}~(2021).

\bibitem{CC22}
Adam Callison and Nicholas Chancellor.
\newblock ``Hybrid quantum-classical algorithms in the noisy intermediate-scale
  quantum era and beyond''.
\newblock \href{https://dx.doi.org/10.1103/PhysRevA.106.010101}{Physical Review
  A {\bf 106}, 010101}~(2022).

\bibitem{MEA20}
Sam McArdle, Suguru Endo, Al\'an Aspuru-Guzik, Simon~C. Benjamin, and Xiao
  Yuan.
\newblock ``Quantum computational chemistry''.
\newblock \href{https://dx.doi.org/10.1103/RevModPhys.92.015003}{Rev. Mod.
  Phys. {\bf 92}, 015003}~(2020).

\bibitem{BBM20}
Bela Bauer, Sergey Bravyi, Mario Motta, and Garnet Kin-Lic Chan.
\newblock ``Quantum algorithms for quantum chemistry and quantum materials
  science''.
\newblock \href{https://dx.doi.org/10.1021/acs.chemrev.9b00829}{Chemical
  Reviews {\bf 120}, 12685--12717}~(2020).

\bibitem{TCC22}
Jules Tilly, Hongxiang Chen, Shuxiang Cao, Dario Picozzi, Kanav Setia, Ying Li,
  Edward Grant, Leonard Wossnig, Ivan Rungger, George~H. Booth, and Jonathan
  Tennyson.
\newblock ``The variational quantum eigensolver: A review of methods and best
  practices''.
\newblock
  \href{https://dx.doi.org/https://doi.org/10.1016/j.physrep.2022.08.003}{Physics
  Reports {\bf 986}, 1--128}~(2022).

\bibitem{MNK18}
K.~Mitarai, M.~Negoro, M.~Kitagawa, and K.~Fujii.
\newblock ``Quantum circuit learning''.
\newblock \href{https://dx.doi.org/10.1103/PhysRevA.98.032309}{Physical Review
  A {\bf 98}, 032309}~(2018).

\bibitem{BLS19}
Marcello Benedetti, Erika Lloyd, Stefan Sack, and Mattia Fiorentini.
\newblock ``Parameterized quantum circuits as machine learning models''.
\newblock \href{https://dx.doi.org/10.1088/2058-9565/ab4eb5}{Quantum Science
  and Technology {\bf 4}, 043001}~(2019).

\bibitem{SK19}
Maria Schuld and Nathan Killoran.
\newblock ``Quantum machine learning in feature hilbert spaces''.
\newblock \href{https://dx.doi.org/10.1103/PhysRevLett.122.040504}{Physical
  Review Letters {\bf 122}, 040504}~(2019).

\bibitem{SBS20}
Maria Schuld, Alex Bocharov, Krysta~M. Svore, and Nathan Wiebe.
\newblock ``Circuit-centric quantum classifiers''.
\newblock \href{https://dx.doi.org/10.1103/PhysRevA.101.032308}{Physical Review
  A {\bf 101}, 032308}~(2020).

\bibitem{CAB21}
Marco Cerezo, Andrew Arrasmith, Ryan Babbush, Simon~C Benjamin, Suguru Endo,
  Keisuke Fujii, Jarrod~R McClean, Kosuke Mitarai, Xiao Yuan, Lukasz Cincio,
  et~al.
\newblock ``Variational quantum algorithms''.
\newblock \href{https://dx.doi.org/10.1038/s42254-021-00348-9}{Nature Reviews
  Physics {\bf 3}, 625--644}~(2021).

\bibitem{BCW01}
Harry Buhrman, Richard Cleve, John Watrous, and Ronald de~Wolf.
\newblock ``Quantum fingerprinting''.
\newblock \href{https://dx.doi.org/10.1103/PhysRevLett.87.167902}{Physical
  Review Letters {\bf 87}, 167902}~(2001).

\bibitem{BBD97}
Adriano Barenco, Andr\'{e} Berthiaume, David Deutsch, Artur Ekert, Richard
  Jozsa, and Chiara Macchiavello.
\newblock ``Stabilization of quantum computations by symmetrization''.
\newblock \href{https://dx.doi.org/10.1137/S0097539796302452}{SIAM Journal on
  Computing {\bf 26}, 1541--1557}~(1997).

\bibitem{KMY03}
Hirotada Kobayashi, Keiji Matsumoto, and Tomoyuki Yamakami.
\newblock ``Quantum {M}erlin-{A}rthur proof systems: Are multiple {M}erlins
  more helpful to {A}rthur?''.
\newblock In Toshihide Ibaraki, Naoki Katoh, and Hirotaka Ono, editors,
  Algorithms and Computation.
\newblock \href{https://dx.doi.org/10.1007/978-3-540-24587-2_21}{Pages
  189--198}.
\newblock Berlin, Heidelberg~(2003). Springer Berlin Heidelberg.

\bibitem{GC13}
Juan~Carlos Garcia-Escartin and Pedro Chamorro-Posada.
\newblock ``{SWAP} test and {H}ong-{O}u-{M}andel effect are equivalent''.
\newblock \href{https://dx.doi.org/10.1103/PhysRevA.87.052330}{Physical Review
  A {\bf 87}, 052330}~(2013).

\bibitem{CSS18}
Lukasz Cincio, Yi\u{g}it Suba\c{s}{\i}, Andrew~T Sornborger, and Patrick~J
  Coles.
\newblock ``Learning the quantum algorithm for state overlap''.
\newblock \href{https://dx.doi.org/10.1088/1367-2630/aae94a}{New Journal of
  Physics {\bf 20}, 113022}~(2018).

\bibitem{War12}
Henry~S. Warren.
\newblock ``Hacker's delight''.
\newblock Addison-Wesley Professional. ~(2012).
\newblock 2nd edition.

\bibitem{Eke21}
Martin Eker{\aa}.
\newblock ``On completely factoring any integer efficiently in a single run of
  an order-finding algorithm''.
\newblock \href{https://dx.doi.org/10.1007/s11128-021-03069-1}{Quantum
  Information Processing {\bf 20}, 1--14}~(2021).

\bibitem{KMT17}
Abhinav Kandala, Antonio Mezzacapo, Kristan Temme, Maika Takita, Markus Brink,
  Jerry~M Chow, and Jay~M Gambetta.
\newblock ``Hardware-efficient variational quantum eigensolver for small
  molecules and quantum magnets''.
\newblock \href{https://dx.doi.org/10.1038/nature23879}{Nature {\bf 549},
  242--246}~(2017).

\bibitem{PVG11}
F.~Pedregosa, G.~Varoquaux, A.~Gramfort, V.~Michel, B.~Thirion, O.~Grisel,
  M.~Blondel, P.~Prettenhofer, R.~Weiss, V.~Dubourg, J.~Vanderplas, A.~Passos,
  D.~Cournapeau, M.~Brucher, M.~Perrot, and E.~Duchesnay.
\newblock ``Scikit-learn: Machine learning in {P}ython''.
\newblock \href{https://www.jmlr.org/papers/volume12/pedregosa11a/pedregosa11a.pdf}{Journal of Machine Learning Research {\bf 12}, 2825--2830}~(2011).

\bibitem{SJG22}
Adam Smith, Bernhard Jobst, Andrew~G. Green, and Frank Pollmann.
\newblock ``Crossing a topological phase transition with a quantum computer''.
\newblock \href{https://dx.doi.org/10.1103/PhysRevResearch.4.L022020}{Physical
  Review Research {\bf 4}, L022020}~(2022).

\bibitem{Llo96}
S.~{Lloyd}.
\newblock ``{Universal Quantum Simulators}''.
\newblock \href{https://dx.doi.org/10.1126/science.273.5278.1073}{Science {\bf
  273}, 1073--1078}~(1996).

\bibitem{BCC15}
Dominic~W. Berry, Andrew~M. Childs, Richard Cleve, Robin Kothari, and
  Rolando~D. Somma.
\newblock ``Simulating hamiltonian dynamics with a truncated taylor series''.
\newblock \href{https://dx.doi.org/10.1103/PhysRevLett.114.090502}{Physical
  Review Letters {\bf 114}, 090502}~(2015).

\bibitem{HHL09}
Aram~W. Harrow, Avinatan Hassidim, and Seth Lloyd.
\newblock ``Quantum algorithm for linear systems of equations''.
\newblock \href{https://dx.doi.org/10.1103/PhysRevLett.103.150502}{Physical
  Review Letters {\bf 103}, 150502}~(2009).

\bibitem{BCC14}
Dominic~W. Berry, Andrew~M. Childs, Richard Cleve, Robin Kothari, and
  Rolando~D. Somma.
\newblock ``Exponential improvement in precision for simulating sparse
  hamiltonians''.
\newblock In Proceedings of the Forty-Sixth Annual ACM Symposium on Theory of
  Computing.
\newblock \href{https://dx.doi.org/10.1145/2591796.2591854}{Page 283–292}.
\newblock STOC '14New York, NY, USA~(2014). Association for Computing
  Machinery.

\bibitem{CBC21}
Laura Clinton, Johannes Bausch, and Toby Cubitt.
\newblock ``Hamiltonian simulation algorithms for near-term quantum hardware''.
\newblock \href{https://dx.doi.org/10.1038/s41467-021-25196-0}{Nature
  communications {\bf 12}, 4989}~(2021).

\bibitem{Pea01}
Karl~Pearson F.R.S.
\newblock ``{LIII.} {O}n lines and planes of closest fit to systems of points
  in space''.
\newblock \href{https://dx.doi.org/10.1080/14786440109462720}{The London,
  Edinburgh, and Dublin Philosophical Magazine and Journal of Science {\bf 2},
  559--572}~(1901).

\bibitem{JC16}
Ian~T. Jolliffe and Jorge Cadima.
\newblock ``Principal component analysis: a review and recent developments''.
\newblock \href{https://dx.doi.org/10.1098/rsta.2015.0202}{Philosophical
  transactions of the Royal Society A: Mathematical, Physical and Engineering
  Sciences {\bf 374}, 20150202}~(2016).

\bibitem{Jol02}
I.T. Jolliffe.
\newblock ``Principal component analysis (2nd ed)''.
\newblock \href{https://dx.doi.org/10.1007/b98835}{Springer}. ~(2002).

\bibitem{LMR14}
Seth Lloyd, Masoud Mohseni, and Patrick Rebentrost.
\newblock ``Quantum principal component analysis''.
\newblock \href{https://dx.doi.org/10.1038/nphys3029}{Nature Physics {\bf 10},
  631--633}~(2014).

\bibitem{Tan21}
Ewin Tang.
\newblock ``Quantum principal component analysis only achieves an exponential
  speedup because of its state preparation assumptions''.
\newblock \href{https://dx.doi.org/10.1103/PhysRevLett.127.060503}{Physical
  Review Letters {\bf 127}, 060503}~(2021).

\bibitem{Koc21}
B\'alint Koczor.
\newblock ``The dominant eigenvector of a noisy quantum state''.
\newblock \href{https://dx.doi.org/10.1088/1367-2630/ac37ae}{New Journal of
  Physics {\bf 23}, 123047}~(2021).

\bibitem{HWB19}
Oscar Higgott, Daochen Wang, and Stephen Brierley.
\newblock ``Variational {Q}uantum {C}omputation of {E}xcited {S}tates''.
\newblock \href{https://dx.doi.org/10.22331/q-2019-07-01-156}{{Quantum} {\bf
  3}, 156}~(2019).

\bibitem{Hau95}
Hermann~A. Haus.
\newblock ``From classical to quantum noise''.
\newblock \href{https://dx.doi.org/10.1364/JOSAB.12.002019}{Journal of the
  Optical Society of America B {\bf 12}, 2019--2036}~(1995).

\bibitem{Poz04}
D.M. Pozar.
\newblock ``Microwave engineering''.
\newblock Wiley. ~(2004).
\newblock Fourth edition.

\bibitem{Cai53}
E.~R. Caianiello.
\newblock ``On quantum field theory --- {I}: Explicit solution of {D}yson's
  equation in electrodynamics without use of {F}eynman graphs''.
\newblock \href{https://dx.doi.org/10.1007/BF02781659}{Il Nuovo Cimento
  (1943-1954) {\bf 10}, 1634--1652}~(1953).

\bibitem{SGL04}
Johannes Skaar, Juan~Carlos {Garc\'{\i}a Escart\'{\i}n}, and Harald Landro.
\newblock ``Quantum mechanical description of linear optics''.
\newblock \href{https://dx.doi.org/10.1119/1.1775241}{American Journal of
  Physics {\bf 72}, 1385--1391}~(2004).

\bibitem{Sch04}
Stefan Scheel.
\newblock ``Permanents in linear optical networks''~(2004).
\newblock
  \href{http://arxiv.org/abs/quant-ph/0406127}{arXiv:quant-ph/0406127}.

\bibitem{AA11}
Scott Aaronson and Alex Arkhipov.
\newblock ``The computational complexity of linear optics''.
\newblock In Proceedings of the 43rd Annual ACM Symposium on Theory of
  Computing.
\newblock \href{https://dx.doi.org/10.1145/1993636.1993682}{Pages 333--342}.
\newblock STOC '11New York, NY, USA~(2011). ACM.

\bibitem{CAK98}
N.J. Cerf, C.~Adami, and P.G. Kwiat.
\newblock ``Optical simulation of quantum logic''.
\newblock \href{https://dx.doi.org/10.1103/PhysRevA.57.R1477}{Physical Review A
  {\bf 57}, 1477--1480}~(1998).

\bibitem{RZB94}
M.~Reck, A.~Zeilinger, H.J. Bernstein, and P.~Bertani.
\newblock ``Experimental realization of any discrete unitary operator''.
\newblock \href{https://dx.doi.org/10.1103/PhysRevLett.73.58}{Physical Review
  Letters {\bf 73}, 58--61}~(1994).

\bibitem{BA14}
Adam Bouland and Scott Aaronson.
\newblock ``Generation of universal linear optics by any beam splitter''.
\newblock \href{https://dx.doi.org/10.1103/PhysRevA.89.062316}{Physical Review
  A {\bf 89}, 062316}~(2014).

\bibitem{Saw16}
A.~{Sawicki}.
\newblock ``{Universality of beamsplitters}''.
\newblock \href{https://dx.doi.org/10.26421/QIC16.3-4-6}{Quantum Information \&
  Computation {\bf 16}, 0291--0312}~(2016).

\bibitem{CHM16}
William~R. Clements, Peter~C. Humphreys, Benjamin~J. Metcalf, W.~Steven
  Kolthammer, and Ian~A. Walmsley.
\newblock ``Optimal design for universal multiport interferometers''.
\newblock \href{https://dx.doi.org/10.1364/OPTICA.3.001460}{Optica {\bf 3},
  1460--1465}~(2016).

\bibitem{GMS18}
Hubert de~Guise, Olivia Di~Matteo, and Luis~L. S\'anchez-Soto.
\newblock ``Simple factorization of unitary transformations''.
\newblock \href{https://dx.doi.org/10.1103/PhysRevA.97.022328}{Physical Review
  A {\bf 97}, 022328}~(2018).

\bibitem{BW21}
B.~A. Bell and I.~A. Walmsley.
\newblock ``{Further compactifying linear optical unitaries}''.
\newblock \href{https://dx.doi.org/10.1063/5.0053421}{APL Photonics {\bf 6},
  070804}~(2021).

\bibitem{CHS15}
Jacques Carolan, Christopher Harrold, Chris Sparrow, Enrique
  Mart{\'\i}n-L{\'o}pez, Nicholas~J. Russell, Joshua~W. Silverstone, Peter~J.
  Shadbolt, Nobuyuki Matsuda, Manabu Oguma, Mikitaka Itoh, Graham~D. Marshall,
  Mark~G. Thompson, Jonathan C.~F. Matthews, Toshikazu Hashimoto, Jeremy~L.
  O{\textquoteright}Brien, and Anthony Laing.
\newblock ``{Universal linear optics}''.
\newblock \href{https://dx.doi.org/10.1126/science.aab3642}{Science {\bf 349}, 711--716}~(2015).

\bibitem{MCS18}
Paolo~L. Mennea, William~R. Clements, Devin~H. Smith, James~C. Gates,
  Benjamin~J. Metcalf, Rex H.~S. Bannerman, Roel Burgwal, Jelmer~J. Renema,
  W.~Steven Kolthammer, Ian~A. Walmsley, and Peter G.~R. Smith.
\newblock ``Modular linear optical circuits''.
\newblock \href{https://dx.doi.org/10.1364/OPTICA.5.001087}{Optica {\bf 5},
  1087--1090}~(2018).

\bibitem{BPC20}
Wim Bogaerts, Daniel P{\'e}rez, Jos{\'e} Capmany, David~AB Miller, Joyce Poon,
  Dirk Englund, Francesco Morichetti, and Andrea Melloni.
\newblock ``Programmable photonic circuits''.
\newblock \href{https://dx.doi.org/10.1038/s41586-020-2764-0}{Nature {\bf 586},
  207--216}~(2020).

\bibitem{EPS20}
Ali~W Elshaari, Wolfram Pernice, Kartik Srinivasan, Oliver Benson, and Val
  Zwiller.
\newblock ``Hybrid integrated quantum photonic circuits''.
\newblock \href{https://dx.doi.org/10.1038/s41566-020-0609-x}{Nature Photonics
  {\bf 14}, 285--298}~(2020).

\bibitem{ABB21}
Juan~M Arrazola, Ville Bergholm, Kamil Br{\'a}dler, Thomas~R Bromley, Matt~J
  Collins, Ish Dhand, Alberto Fumagalli, Thomas Gerrits, Andrey Goussev,
  Lukas~G Helt, et~al.
\newblock ``Quantum circuits with many photons on a programmable nanophotonic
  chip''.
\newblock \href{https://dx.doi.org/10.1038/s41586-021-03202-1}{Nature {\bf
  591}, 54--60}~(2021).

\bibitem{TMS21}
Caterina Taballione, Reinier van~der Meer, Henk~J Snijders, Peter Hooijschuur,
  Jörn~P Epping, Michiel de~Goede, Ben Kassenberg, Pim Venderbosch, Chris
  Toebes, Hans van~den Vlekkert, Pepijn W~H Pinkse, and Jelmer~J Renema.
\newblock ``A universal fully reconfigurable 12-mode quantum photonic
  processor''.
\newblock \href{https://dx.doi.org/10.1088/2633-4356/ac168c}{Materials for
  Quantum Technology {\bf 1}, 035002}~(2021).

\bibitem{HPG22}
Francesco Hoch, Simone Piacentini, Taira Giordani, Zhen-Nan Tian, Mariagrazia
  Iuliano, Chiara Esposito, Anita Camillini, Gonzalo Carvacho, Francesco
  Ceccarelli, Nicol{\'o} Spagnolo, Andrea Crespi, Fabio Sciarrino, and Roberto
  Osellame.
\newblock ``Reconfigurable continuously-coupled 3{D} photonic circuit for boson
  sampling experiments''.
\newblock \href{https://dx.doi.org/10.1038/s41534-022-00568-6}{npj Quantum
  Information {\bf 8}, 1--7}~(2022).

\bibitem{PMS14}
Alberto Peruzzo, Jarrod McClean, Peter Shadbolt, Man-Hong Yung, Xiao-Qi Zhou,
  Peter~J Love, Al{\'a}n Aspuru-Guzik, and Jeremy~L O’brien.
\newblock ``A variational eigenvalue solver on a photonic quantum processor''.
\newblock \href{https://dx.doi.org/10.1038/ncomms5213}{Nature communications
  {\bf 5}, 1--7}~(2014).

\bibitem{SvE11}
Lucia Schwarz and S.~J. van Enk.
\newblock ``Detecting the drift of quantum sources: Not the de {F}inetti
  theorem''.
\newblock \href{https://dx.doi.org/10.1103/PhysRevLett.106.180501}{Physical
  Review Letters {\bf 106}, 180501}~(2011).

\bibitem{ABS92}
L.~Allen, M.~W. Beijersbergen, R.~J.~C. Spreeuw, and J.~P. Woerdman.
\newblock ``Orbital angular momentum of light and the transformation of
  laguerre-gaussian laser modes''.
\newblock \href{https://dx.doi.org/10.1103/PhysRevA.45.8185}{Physical Review A
  {\bf 45}, 8185--8189}~(1992).

\bibitem{ABP03}
L.~Allen, S.M. Barnett, and M.J. Padgett.
\newblock ``{Optical Angular Momentum}''.
\newblock \href{https://dx.doi.org/10.1201/9781482269017}{Institute of Physics
  Publishing, Bristol, UK}. ~(2003).

\bibitem{WPS21}
Alan~E. Willner, Kai Pang, Hao Song, Kaiheng Zou, and Huibin Zhou.
\newblock ``{Orbital angular momentum of light for communications}''.
\newblock \href{https://dx.doi.org/10.1063/5.0054885}{Applied Physics Reviews
  {\bf 8}, 041312}~(2021).

\bibitem{MHH97}
A.~Muller, T.~Herzog, B.~Huttner, W.~Tittel, H.~Zbinden, and N.~Gisin.
\newblock ``{`Plug and play' systems for quantum cryptography}''.
\newblock \href{https://dx.doi.org/10.1063/1.118224}{Applied Physics Letters
  {\bf 70}, 793--795}~(1997).

\bibitem{WLP21}
Robert~Ian Woodward, YS~Lo, M~Pittaluga, M~Minder, TK~Para{\"\i}so,
  M~Lucamarini, ZL~Yuan, and AJ~Shields.
\newblock ``Gigahertz measurement-device-independent quantum key distribution
  using directly modulated lasers''.
\newblock \href{https://dx.doi.org/10.1038/s41534-021-00394-2}{npj Quantum
  Information {\bf 7}, 58}~(2021).

\bibitem{SBC09}
Valerio Scarani, Helle Bechmann-Pasquinucci, Nicolas~J. Cerf, Miloslav
  Du\ifmmode~\check{s}\else \v{s}\fi{}ek, Norbert L\"utkenhaus, and Momtchil
  Peev.
\newblock ``The security of practical quantum key distribution''.
\newblock \href{https://dx.doi.org/10.1103/RevModPhys.81.1301}{Reviews of
  Modern Physics {\bf 81}, 1301--1350}~(2009).

\bibitem{XMZ20}
Feihu Xu, Xiongfeng Ma, Qiang Zhang, Hoi-Kwong Lo, and Jian-Wei Pan.
\newblock ``Secure quantum key distribution with realistic devices''.
\newblock \href{https://dx.doi.org/10.1103/RevModPhys.92.025002}{Reviews of
  Modern Physics {\bf 92}, 025002}~(2020).


\bibitem{SG21}
D.~ Stilck~Fran{\c{c}}a and R.~ Garc\'ia-Patron.
\newblock Limitations of optimization algorithms on noisy quantum devices.
\newblock  \href{https://doi.org/10.1038/s41567-021-01356-3}{{\em Nature Physics}, 17(11):1221--1227}, 2021.

\bibitem{WFC21}
S.~ Wang, E.~ Fontana, M.~ Cerezo, K.~ Sharma, A.~ Sone, L.~ Cincio, and P.~J.~ Coles.
\newblock Noise-induced barren plateaus in variational quantum algorithms.
\newblock \href{https://doi.org/10.1038/s41467-021-27045-6}{{\em Nature communications}, 12(1):6961}, 2021.

\bibitem{KLP19}
S.~ Khatri, R.~ LaRose, A.~ Poremba, L.~ Cincio, A.~T.~ Sornborger, and P.~J.~ Coles.
\newblock Quantum-assisted quantum compiling.
\newblock \href{https://doi.org/10.22331/q-2019-05-13-140}{{\em Quantum}, 3:140}, 2019.

\bibitem{KMvB19}
C.~ Kokail, C.~ Maier, R.~ van Bijnen, T.~ Brydges, M.~K.~ Joshi, P.~ Jurcevic, C.~A.~ Muschik, P.~ Silvi, R.~ Blatt, C.~F.~ Roos, and P.~ Zoller.
\newblock Self-verifying variational quantum simulation of lattice models.
\newblock \href{https://doi.org/10.1038/s41586-019-1177-4}{{\em Nature}, 569(7756):355--360}, 2019.

\bibitem{KUY20}
A.~ Kardashin, A.~ Uvarov, D.~ Yudin, and J.~ Biamonte.
\newblock Certified variational quantum algorithms for eigenstate preparation.
\newblock \href{https://doi.org/10.1103/PhysRevA.102.052610}{{\em Physical Review A}, 102:052610}, 2020.

\bibitem{LTO19}
R.~ LaRose, A.~ Tikku, {\'E}.~ O’Neel-Judy, L.~ Cincio, and P.~J.~ Coles.
\newblock Variational quantum state diagonalization.
\newblock \href{https://doi.org/10.1038/s41534-019-0167-6}{{\em npj Quantum Information}, 5(1):57}, 2019.

\bibitem{CSA22}
M.~ Cerezo, K.~ Sharma, A.~ Arrasmith, and P.~J.~ Coles.
\newblock Variational quantum state eigensolver.
\newblock \href{https://doi.org/10.1038/s41534-022-00611-6}{{\em npj Quantum Information}, 8(1):113}, 2022.

\bibitem{Fra61}
J.~G.~F. Francis.
\newblock ``{The QR Transformation A Unitary Analogue to the LR
  Transformation—Part 1}''.
\newblock \href{https://dx.doi.org/10.1093/comjnl/4.3.265}{The Computer Journal
  {\bf 4}, 265--271}~(1961).

\bibitem{Fra62}
J.~G.~F. Francis.
\newblock ``{The QR Transformation—Part 2}''.
\newblock \href{https://dx.doi.org/10.1093/comjnl/4.4.332}{The Computer Journal
  {\bf 4}, 332--345}~(1962).

\bibitem{Kub62}
V.N. Kublanovskaya.
\newblock ``On some algorithms for the solution of the complete eigenvalue
  problem''.
\newblock
  \href{https://dx.doi.org/https://doi.org/10.1016/0041-5553(63)90168-X}{USSR
  Computational Mathematics and Mathematical Physics {\bf 1}, 637--657}~(1962).

\bibitem{ABB99}
E.~Anderson, Z.~Bai, C.~Bischof, S.~Blackford, J.~Demmel, J.~Dongarra,
  J.~Du~Croz, A.~Greenbaum, S.~Hammarling, A.~McKenney, and D.~Sorensen.
\newblock ``{LAPACK} users' guide''.
\newblock \href{https://dx.doi.org/10.1137/1.9780898719604}{Society for
  Industrial and Applied Mathematics}. Philadelphia, PA~(1999).
\newblock Third edition.

\bibitem{Dem97}
James~W Demmel.
\newblock ``Applied numerical linear algebra''.
\newblock SIAM. ~(1997).

\bibitem{Ste01}
Gilbert~W Stewart.
\newblock ``Matrix algorithms: {V}olume {II}: {E}igensystems''.
\newblock SIAM. ~(2001).

\bibitem{Bjo15}
{\AA}ke Bj{\"o}rck.
\newblock ``Numerical methods in matrix computations''.
\newblock \href{https://dx.doi.org//10.1007/978-3-319-05089-8}{Springer}.
  ~(2015).

\bibitem{Kit95}
A~Yu Kitaev. \newblock ``Quantum measurements and the abelian stabilizer problem''~(1995).
\newblock  \href{http://arxiv.org/abs/quant-ph/9}{arXiv:quant-ph/951102}.

\bibitem{AL99}
Daniel~S. Abrams and Seth Lloyd.
\newblock ``Quantum algorithm providing exponential speed increase for finding
  eigenvalues and eigenvectors''.
\newblock \href{https://dx.doi.org/10.1103/PhysRevLett.83.5162}{Physical Review
  Letters {\bf 83}, 5162--5165}~(1999).

\bibitem{CWV20}
Shuxiang Cao, Leonard Wossnig, Brian Vlastakis, Peter Leek, and Edward Grant.
\newblock ``Cost-function embedding and dataset encoding for machine learning
  with parametrized quantum circuits''.
\newblock \href{https://dx.doi.org/10.1103/PhysRevA.101.052309}{Physical Review
  A {\bf 101}, 052309}~(2020).

\bibitem{AJL06}
Dorit Aharonov, Vaughan Jones, and Zeph Landau.
\newblock ``A polynomial quantum algorithm for approximating the {J}ones
  polynomial''.
\newblock In Proceedings of the Thirty-Eighth Annual ACM Symposium on Theory of
  Computing.
\newblock \href{https://dx.doi.org/10.1145/1132516.1132579}{Page 427–436}.
\newblock STOC '06New York, NY, USA~(2006). Association for Computing
  Machinery.

\bibitem{PMR09}
G.~Passante, O.~Moussa, C.~A. Ryan, and R.~Laflamme.
\newblock ``Experimental approximation of the {J}ones polynomial with one
  quantum bit''.
\newblock \href{https://dx.doi.org/10.1103/PhysRevLett.103.250501}{Physical
  Review Letters {\bf 103}, 250501}~(2009).

\bibitem{AFC14}
Mateus Ara\'ujo, Adrien Feix, Fabio Costa, and \v{C}aslav Brukner.
\newblock ``Quantum circuits cannot control unknown operations''.
\newblock \href{https://dx.doi.org/10.1088/1367-2630/16/9/093026}{New Journal
  of Physics {\bf 16}, 093026}~(2014).

\bibitem{LLS08}
Yang Liu, Gui~Lu Long, and Yang Sun.
\newblock ``Analytic one-bit and {CNOT} gate constructions of general n-qubit
  controlled gates''.
\newblock \href{https://dx.doi.org/10.1142/S0219749908003621C}{International
  Journal of Quantum Information {\bf 6}, 447--462}~(2008).

\bibitem{MF19}
Kosuke Mitarai and Keisuke Fujii.
\newblock ``Methodology for replacing indirect measurements with direct
  measurements''.
\newblock \href{https://dx.doi.org/10.1103/PhysRevResearch.1.013006}{Physical
  Review Research {\bf 1}, 013006}~(2019).

\bibitem{AA21}
Guillermo Alonso-Linaje and Parfait Atchade-Adelomou.
\newblock ``{EVA}: a quantum exponential value approximation
  algorithm''~(2021).
\newblock
  \href{https://doi.org/10.48550/arXiv.2106.08731}{arXiv:2106.08731v1}.


\bibitem{WRZ21}
Bujiao Wu, Maharshi Ray, Liming Zhao, Xiaoming Sun, and Patrick Rebentrost.
\newblock ``Quantum-classical algorithms for skewed linear systems with an
  optimized {H}adamard test''.
\newblock \href{https://dx.doi.org/10.1103/PhysRevA.103.042422}{Physical Review
  A {\bf 103}, 042422}~(2021).

\bibitem{HSC22}
Zo\"e Holmes, Kunal Sharma, M.~Cerezo, and Patrick~J. Coles.
\newblock ``Connecting ansatz expressibility to gradient magnitudes and barren
  plateaus''.
\newblock \href{https://dx.doi.org/10.1103/PRXQuantum.3.010313}{PRX Quantum
  {\bf 3}, 010313}~(2022).

\bibitem{DTY22}
Yuxuan Du, Zhuozhuo Tu, Xiao Yuan, and Dacheng Tao.
\newblock ``Efficient measure for the expressivity of variational quantum
  algorithms''.
\newblock \href{https://dx.doi.org/10.1103/PhysRevLett.128.080506}{Physical
  Review Letters {\bf 128}, 080506}~(2022).

\bibitem{FHJ22}
Lena Funcke, Tobias Hartung, Karl Jansen, Stefan K{\"{u}}hn, and Paolo
  Stornati.
\newblock ``Dimensional {E}xpressivity {A}nalysis of {P}arametric {Q}uantum
  {C}ircuits''.
\newblock \href{https://dx.doi.org/10.22331/q-2021-03-29-422}{{Quantum} {\bf
  5}, 422}~(2021).

\bibitem{Gar23}
Juan~Carlos Garcia-Escartin.
\newblock ``Example code for the algorithm''.
\newblock \url{https://github.tel.uva.es/juagar/QVEF}~(2023).
\end{thebibliography}
\section*{References}

\newcommand{\noopsort}[1]{} \newcommand{\printfirst}[2]{#1}
  \newcommand{\singleletter}[1]{#1} \newcommand{\switchargs}[2]{#2#1}

\end{document}